%% file: main.tex
\documentclass[sigplan,screen]{acmart}
\AtBeginDocument{%
  }

\usepackage[]{hyperref}
\usepackage{algorithmic}
\usepackage{graphicx}
\usepackage{textcomp}
\usepackage{xcolor}
\usepackage{makecell}
\usepackage{diagbox}
\usepackage[normalem]{ulem}
\usepackage{listings}
\usepackage{subfig}
\usepackage{multirow}
\usepackage{multicol}
\usepackage{booktabs} 
\usepackage[ruled,vlined,linesnumbered]{algorithm2e}
\usepackage[most]{tcolorbox}
\usepackage{cleveref}

\copyrightyear{2025}
\acmYear{2025}
\setcopyright{acmlicensed}\acmConference[ASPLOS '25]{Proceedings of the 30th ACM
International Conference on Architectural Support for Programming Languages and
Operating Systems, Volume 3}{March 30-April 3, 2025}{Rotterdam, Netherlands}
\acmBooktitle{Proceedings of the 30th ACM International Conference on Architectural
Support for Programming Languages and Operating Systems, Volume 3 (ASPLOS '25),
March 30-April 3, 2025, Rotterdam, Netherlands}
\acmDOI{10.1145/3676642.3736119}
\acmISBN{979-8-4007-1080-3/2025/03}

\lstdefinestyle{code_style}{
  language=bash,
  basicstyle=\footnotesize\ttfamily,
  numbers=none,
  keywordstyle=\color{black},
  numberstyle=\tiny,
  numbersep=4pt,
  frame=tblr,
  columns=fullflexible,
  backgroundcolor=\color{lightgray!20},
  linewidth=0.95\linewidth,
  xleftmargin=0.03\linewidth
}
\makeatletter
\newcommand{\algorithmfootnote}[2][\small]{%
  \let\old@algocf@finish\@algocf@finish
  \def\@algocf@finish{\old@algocf@finish
    \leavevmode\rlap{\begin{minipage}{\linewidth}
    #1#2
    \end{minipage}}
  }%
}

\newenvironment{CompactItemize}%
{\begin{list}{$\bullet$}%
    {\leftmargin=\parindent \itemsep=2pt \topsep=2pt
    \parsep=0pt \partopsep=0pt}}%
{\end{list}}

\newcommand{\circledtext}[1]{\raisebox{.5pt}{\textcircled{\bf\raisebox{-0.8pt} {\small #1}}}}

\lstdefinestyle{BashInputStyle}{
  language=bash,
  basicstyle=\footnotesize\ttfamily,
  numbers=none,
  keywordstyle=\color{black},
  numberstyle=\tiny,
  numbersep=4pt,
  frame=tblr,
  columns=fullflexible,
  backgroundcolor=\color{lightgray!20},
  linewidth=0.95\linewidth,
  xleftmargin=0.03\linewidth,
  breaklines=true
}

\def\projectname{HybridTier}

\begin{document}

\title{\projectname{}: an Adaptive and Lightweight CXL-Memory Tiering System}

\author{Kevin Song}
\affiliation{%
  \institution{University of Toronto, Vector Institute}
  \city{Toronto}
  \country{Canada}
}
\email{xinyang.song@utoronto.ca}

\author{Jiacheng Yang}
\affiliation{%
  \institution{University of Toronto, Vector Institute}
  \city{Toronto}
  \country{Canada}
}
\email{jiacheng.yang@mail.utoronto.ca}

\author{Zixuan Wang}
\affiliation{%
  \institution{University of California San Diego}
  \city{San Diego}
  \country{USA}
}
\email{zxwang@ucsd.edu}

\author{Jishen Zhao}
\affiliation{%
  \institution{University of California San Diego}
  \city{San Diego}
  \country{USA}
}
\email{jzhao@ucsd.edu}

\author{Sihang Liu}
\affiliation{%
  \institution{University of Waterloo}
  \city{Waterloo}
  \country{Canada}
}
\email{sihangliu@uwaterloo.ca}

\author{Gennady Pekhimenko}
\affiliation{%
  \institution{University of Toronto, Vector Institute}
  \city{Toronto}
  \country{Canada}
}
\email{pekhimenko@cs.toronto.edu}

\renewcommand{\shortauthors}{Song et al.}

\begin{abstract}
Modern workloads are demanding increasingly larger memory capacity.
Compute Express Link (CXL)-based memory tiering has emerged as a promising solution for addressing this problem by utilizing traditional DRAM alongside slow-tier CXL memory devices.
We analyze prior tiering systems and observe two challenges for high-performance memory tiering: adapting to skewed but dynamically varying data hotness distributions while minimizing memory and cache overhead due to tiering. 

To address these challenges, we propose \projectname{}, an adaptive and lightweight tiering system for CXL memory.
\projectname{} tracks both long-term data access frequency and short-term access momentum \emph{simultaneously} to accurately capture and adapt to shifting hotness distributions.
\projectname{} reduces the metadata memory overhead by tracking data accesses \emph{probabilistically}, obtaining higher memory efficiency by trading off a small amount of tracking inaccuracy that has a negligible impact on application performance.
To reduce cache overhead, \projectname{} uses lightweight data structures that optimize for data locality to track data hotness.
Our evaluations show that \projectname{} outperforms prior systems by up to $91\%$ ($19\%$ geomean), incurring $2.0-7.8\times$ less memory overhead and $1.7-3.5\times$ less cache misses.
\projectname{} is open source at \url{https://github.com/kevins981/hybridtier-asplos25-artifact}.
\end{abstract}



\maketitle

\input{sec/01_introduction}
\input{sec/02_background}
\input{sec/03_overview} 
\input{sec/04a_design}

\input{sec/04b_implementation}
\input{sec/05_methodology}

\input{sec/06_evaluation}
\input{sec/07_discussion}
\input{sec/08_related_works}

\input{sec/09_conclusion}


\bibliographystyle{plain}
\bibliography{sample-base}

\appendix
\clearpage
\input{sec/appendix}
\end{document}

%% file: sec/01_introduction.tex
\section{Introduction}

Modern applications~\cite{cachelib_paper, twitter_paper, gap, xgboost_git} demand increasingly large memory capacity and high bandwidth.
To keep up with this fast growth, one potential solution is to install more and higher-capacity DRAM modules. 
However, the amount of DRAM within a single server is limited due to space constraints.
In addition, main memory is already one of the most expensive components in data center servers. 
Meta reports that 37$\%$ of the total cost of ownership per rack is spent on memory alone~\cite{tpp}. 
Purchasing more DRAM modules will only exacerbate this trend. 
Moreover, the growth in DRAM density has been slowing down since the last decade~\cite{mem_prospect, mem_scaling_onur}, further limiting the capacity scalability of main memory. 

\begin{figure}[t]
\centering
\includegraphics[width=.8\linewidth]{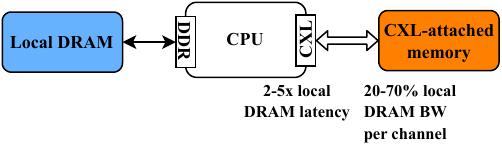}
\caption{Illustration for CXL-attached memory expansion. Performance numbers obtained from~\cite{tpp, uiuc_cxl}.}
\label{fig:cxl_block}
\Description{}
\end{figure}

Compute Express Link (CXL) based memory tiering is a promising solution to this challenge. 
CXL is an industry-standard interconnect protocol \cite{cxl}.
Memory (e.g., DDR4/5 DRAM) attached through CXL interface is byte-addressable, directly accessible by the host CPU, and supports standard memory allocation interfaces. 
Compared to local DRAM, CXL-attached memory has a larger capacity and lower cost-per-GB, as the CXL bus consumes less power and allows DRAM modules to be utilized in a more compact form factor.
On the other hand, CXL memory suffers from higher latency and lower bandwidth.
As shown in Figure~\ref{fig:cxl_block}, compared to local DRAM, CXL memory devices introduce 50$-$100 ns of additional access latency~\cite{tpp}, while achieving 20$-$70$\%$ of its bandwidth~\cite{uiuc_cxl}.
Therefore, a tiering system should prioritize placing hot data in local DRAM (fast-tier) while keeping cold data in CXL memory (slow-tier) for better performance.

However, achieving high tiering performance is challenging. 
We profile large memory workloads and make two observations.
1) Real-world workloads often exhibit skewed yet dynamically varying data hotness distributions \cite{daemon, twitter_paper, gap}.
For instance, in-memory caches often exhibit Zipf or power-law access distributions \cite{cachelib_paper, twitter_paper}, where the majority of accesses are focused on a small fraction of data.
At the same time, hot data can become cold in a matter of minutes~\cite{cachelib_paper, twitter_paper}, causing changes in data hotness distribution.
2) Managing data access statistics can incur significant overhead.
Large memory servers with terabytes of memory contains billions of pages. 
Tiering metadata associated with each page combined can decrease the cost-effectiveness of tiering.
In addition, frequent access of tiering metadata generates high amounts of CPU cache traffic, degrading application performance due to resource contentions. 
Therefore, we argue that an ideal tiering system should satisfy three requirements: 1) accurately capture the hot set by placing the hottest data in fast-tier memory 2) quickly adapt to changes in the hotness distribution 3) minimize tiering metadata overhead.

However, prior systems do not meet all three requirements. 
One class of work adopts frequency-based tiering \cite{memtis, hemem, flexmem} by storing the access counts of each page to build a hotness histogram.
Based on this histogram, the tiering system places the hottest pages in the fast-tier, satisfying requirement 1.
However, frequency-based systems do not meet requirements 2 and 3. 
To maintain histogram freshness, prior works periodically perform ``cooling'' by reducing the page access counter values for all pages. 
In Section \ref{sec:varying_hotness}, we extensively analyze this type of freshness mechanism and demonstrate how it causes tiering systems to be slow at adapting to hotness changes, failing requirement 2.
Furthermore, to store access counts for billions of pages, prior frequency-based systems can incur gigabytes of memory overhead per server, translating to lower cost-effectiveness. 
At the same time, data structures used by prior works to organize a large number of access counters lack data locality.
Since access counters need to be frequently updated, tiering systems can generate a large number of cache misses and degrade application performance, failing requirement 3.

Another class of prior works \cite{autonuma_gap, tpp, multiclock} is recency-based tiering, which uses access recency to approximate data hotness. 
Such systems use metrics such as time between consecutive page faults to make data placement decisions.
In general, recency-based systems meet requirement 3 since recency statistics are less resource-intensive to manage than frequency counterparts \cite{memtis, hint_fault_patch, autonuma_huang}.
Intuitively, recency systems only need to store the most recent event without the need to track historical events.
However, recency-based systems fail to satisfy requirement 1.
Since such systems only consider access statics in a short time interval, they are susceptible to misclassifying cold pages as hot \cite{ying_comment, memtis, multiclock}.
Recency-based systems also do not satisfy requirement 2.
While recency metrics are naturally ``fresh'' and thus do not require cooling, this alone is not sufficient to achieve high adaptiveness.
We demonstrate in Section \ref{sec:varying_hotness} that satisfying requirement 1 is a prerequisite for requirement 2.

In this work, we propose \projectname{}, an application transparent tiering system that addresses the three requirements.
To manage workloads with skewed but dynamically varying hotness distributions, \projectname{} maintains two metrics for each page to capture both long-term access history and short-term hotness variations.
\projectname{} considers statistics from both metrics to enable a flexible migration policy.
For promotion--moving data from the slow-tier to fast-tier--\projectname{} promotes not only pages with high historical hotness but also pages with high access momentum in the short-term to quickly identify pages that have recently become hot.
For demotion, \projectname{} adopts a second-chance policy to swiftly demote pages that were historically hot but have recently turned cold. 

However, maintaining two metrics for every page can exacerbate the problem of high metadata memory overhead. 
To address this challenge, we make the observation that tiering systems can tolerate a small amount of tracking inaccuracy without noticeably affecting application performance.
Based on this observation, \projectname{} tracks memory accesses using counting bloom filters (CBF) \cite{cbf_orig}, a \textit{probabilistically} data structure that trades higher memory efficiency for lower tracking accuracy.
\projectname{}'s CBF is also cache efficient, as it is more compact and requires fewer memory accesses compared to data structures used by prior works.  
To further reduce cache misses, \projectname{} optimizes the locality of CBF by adopting blocked CBF \cite{bcbf, caffeine_bcbf}.

In summary, we make the following contributions:
\begin{CompactItemize}
  \item We analyze existing tiering systems and reveal three new findings: 1) Adapting to changing hotness is challenging, causing suboptimal performance under real-world workloads  2) Maintaining historical access information can incur high metadata memory overhead 3) Tiering systems can suffer from high number of cache misses due to poor data locality during metadata updates. 
  \item We introduce \mbox{\projectname{}}, an application transparent memory tiering system that is adaptive and lightweight. \projectname{} adopts a novel access tracking method that captures both long-term hotness distribution and short-term changes in data hotness. At the same time, \projectname{} significantly reduces metadata memory consumption and cache misses by adopting probabilistic access tracking and locality-optimized data structures.
  \item We compare the performance of \projectname{} against three state-of-the-art tiering systems over six large memory workloads while varying the fast-to-slow tier memory ratio. \projectname{} outperforms prior works by an average of 19\% while incurring $2.0-7.8\times$ less memory overhead and $1.7-3.5\times$ less cache misses due to tiering.
\end{CompactItemize}

%% file: sec/02_background.tex
\section{Background and Motivation}

\subsection{CXL-Enabled Memory Tiering Systems}
%
CXL is an open industry standard interconnect running on top of the PCIe physical layer~\cite{cxl_whitepaper1, cxl_micron, cxl_samsung, cxl_pan}.
The key goal of CXL is to better support heterogeneous computing and memory capabilities in future data center architectures.
%
%
Since its introduction in 2019, the CXL ecosystem has been under rapid development.
With support from major vendors across the data center stack, CXL is widely believed to make a significant impact in future data center architectures.

A CXL-enabled memory tiering system utilizes both regular CPU-attached DRAM (fast-tier) and CXL-attached memory (slow-tier) in the same system.
Local DRAM offers better latency and bandwidth but has a higher cost-per-GB, while CXL memory provides higher capacity but suffers from lower access performance.
Therefore, the goal of a tiered memory system is to accurately identify hot and cold data and place them into the local DRAM and CXL-memory, respectively.

\begin{figure}[t]
  \centering
  \subfloat[Graph analytic: Page Rank.]{\includegraphics[width=0.247\textwidth]{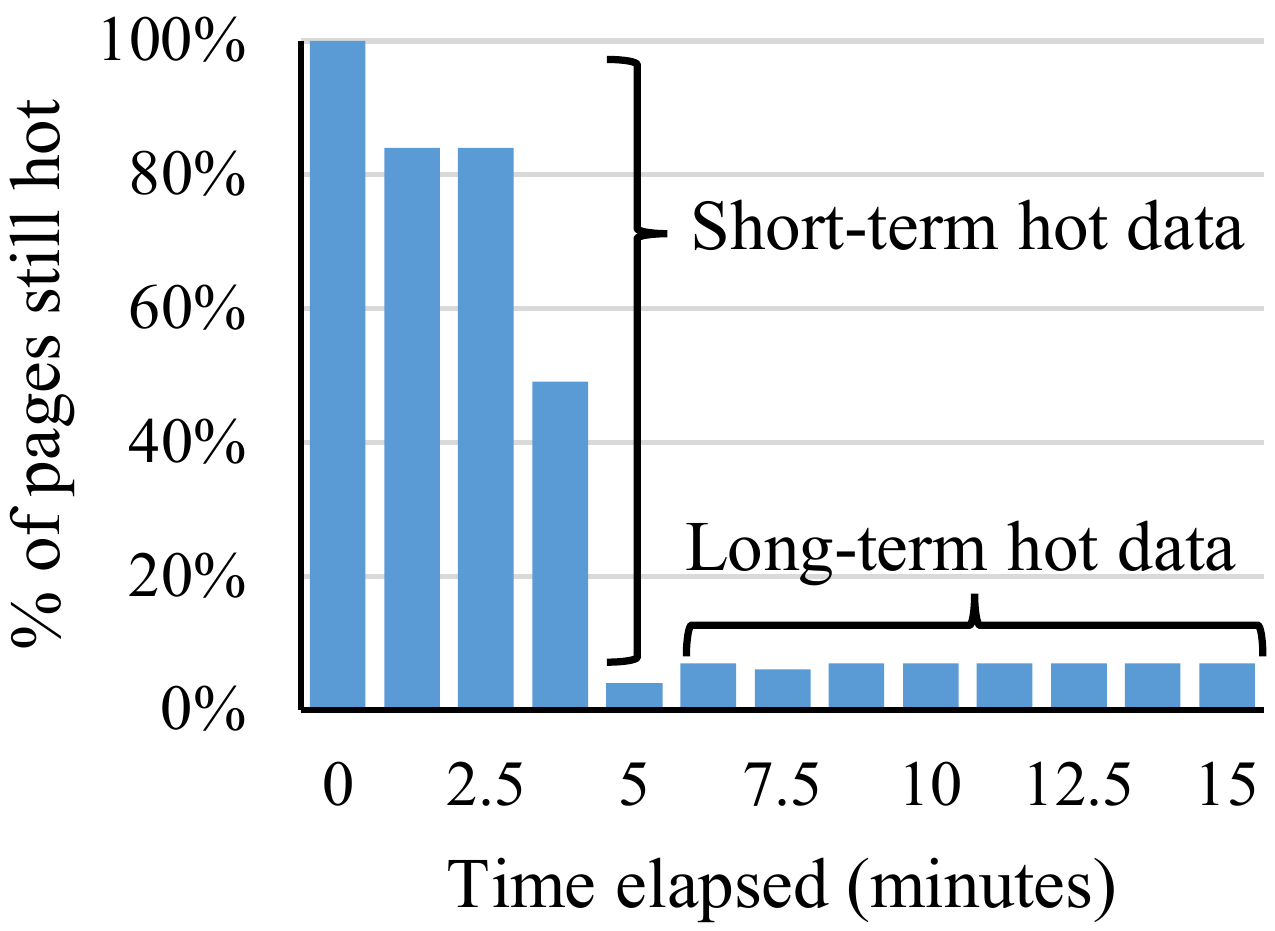}}
\hspace{0.3mm}
  \subfloat[Machine learning: XGBoost.]{\includegraphics[width=0.218\textwidth]{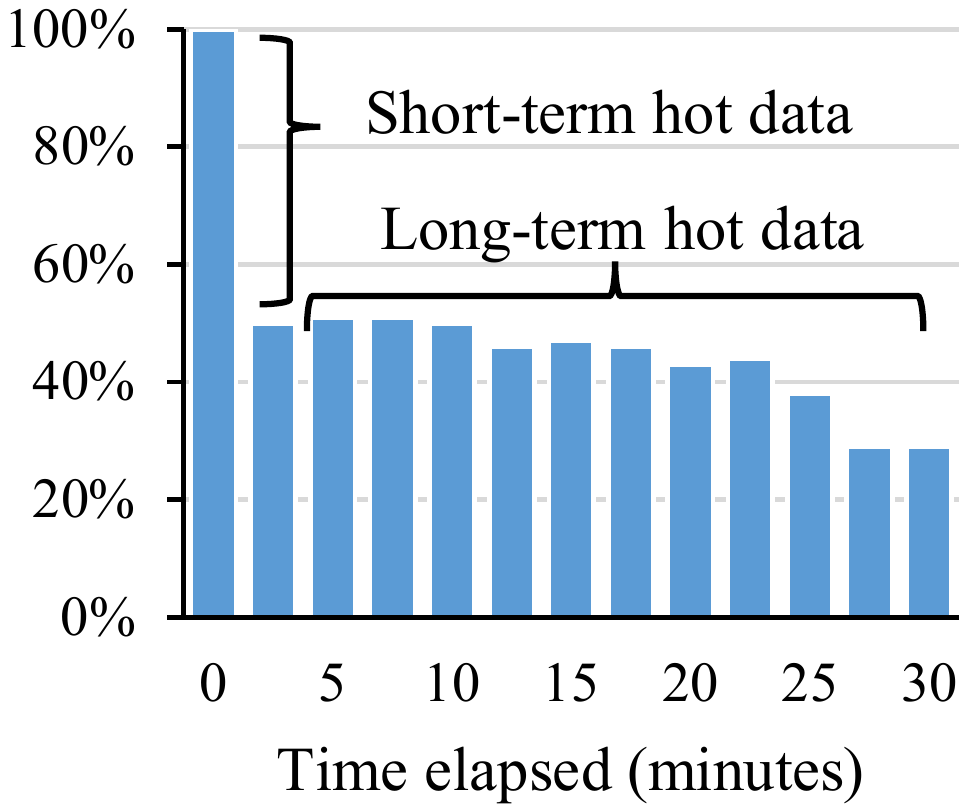}}
  \caption{Data hotness distribution changes rapidly. The Y-axis is the fraction of pages that were hot at time 0 and remained hot over a certain time (X-axis). In both workloads, most pages are no longer hot after just 5 minutes. }
\label{fig:churn}
\Description{}
\end{figure}

\subsection{Dynamic Data Hotness Distribution} \label{sec:background_dynamic}

Prior studies have shown that many large memory workloads exhibit skewed memory access distribution.
Twitter and Meta both report that within a 24-hour window, the popularity of in-memory caching workloads largely follows the Zipfian distribution with a high degree of skewness~\cite{twitter_paper, cachelib_paper, cachelib_ssd}.
For instance, approximately 80\% of accesses to Meta's object storage cache focus on the top 10\% most popular items.
%
%

At the same time, large memory workloads also tend to have dynamically changing hotness distributions.
Meta~\cite{cachelib_paper} report that production in-memory caches experience rapidly shifting access distributions.
At any moment in time, 50\% of popular objects are no longer popular after just 10 minutes.
%
%
Twitter~\cite{twitter_data, twitter_paper} reported that production in-memory caches often use short time-to-live (TTL) in the order of minutes, where objects are removed from the cache after the TTL expires.
%
%
\autoref{fig:churn} shows that throughput-oriented workloads such as graph analytic~\cite{gap} and machine learning training~\cite{xgboost_git} also experience dynamic access distributions.
In Page Rank and XGBoost, over 90\% and 50\% of initially hot pages are no longer hot after just 5 minutes. 

\subsection{Prior Tiering Systems}

This section analyzes prior works based on three requirements of an effective tiering system: (1) ability to accurately capture hot data (2) adaptability to changing hotness distributions (3) tiering metadata overhead.

\subsubsection{Accurately Capture Hot Data} \label{sec:capture_hot}
An essential goal of tiering systems is to maximize memory performance by placing hot data in fast-tier memory.
Therefore, accurately identifying hot data is critical.
Since different applications can exhibit different levels of skewness in their hotness distributions, an effective tiering system must place only the hottest pages in fast-tier memory to maximize performance.

Fortunately, recent works can satisfy this requirement effectively.
For instance, Memtis \cite{memtis}, a state-of-the-art tiering system, maintains a histogram to track the overall access frequency distribution of memory pages.
By understanding the overall hotness distribution and the fast-tier memory capacity, Memtis can accurately calculate the hotness threshold to ensure only the hottest data are placed in the fast-tier. 
%

\subsubsection{Adapting to Varying Hotness Distributions} \label{sec:varying_hotness}

As discussed in \autoref{sec:background_dynamic}, real-world workloads often exhibit dynamically changing hotness distributions.
As a result, in addition to accurately capturing hot data, tiering systems should also \textit{quickly} identify pages that are turning hot and turning cold.
Without this ability, pages that are no longer hot would be left in fast-tier memory, consuming precious resources and leaving performance on the table. 
To analyze the adaptability of tiering systems, we categorize prior works into frequency-based and recency-based according to the hotness metric used. 

\textbf{Frequency-based Systems. }
To identify hot pages, one line of work, such as Memtis \cite{memtis}, tracks the overall data access distribution by maintaining dedicated frequency counters for each page. 
A page is considered hot if its accumulated access frequency exceeds a hotness threshold. 
To ensure freshness, frequency-based systems typically implement exponential moving average (EMA) \cite{memtis, hemem} with a decay factor of 2, where page access counters are periodically cooled by dividing all access counters by two\footnote{Decay factor 2 is typically used since it can be implemented using bit shift.}. 
The cooling period $C$ is a pre-determined parameter.  
%
%
%


\begin{figure}[t]
  \centering
  \subfloat[Bottom series: a page accessed 50 times per minute for 10 minutes. Top series: EMA score of this page.]{\includegraphics[width=0.22\textwidth]{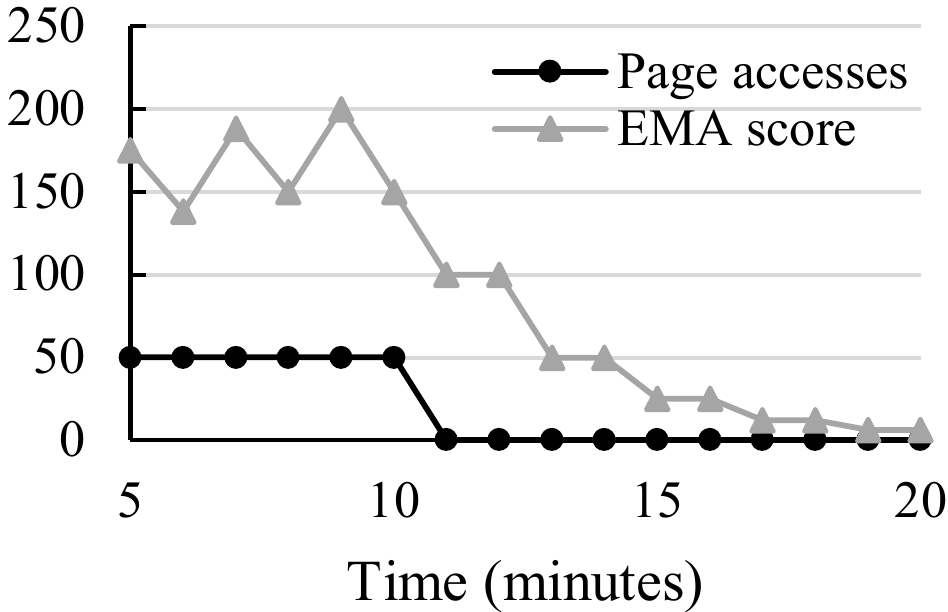}}
\hspace{1mm}
  \subfloat[Fraction of cold, warm, and hot pages in CacheLib workload identified by EMA score under different cooling period $C$.]{\includegraphics[width=0.245\textwidth]{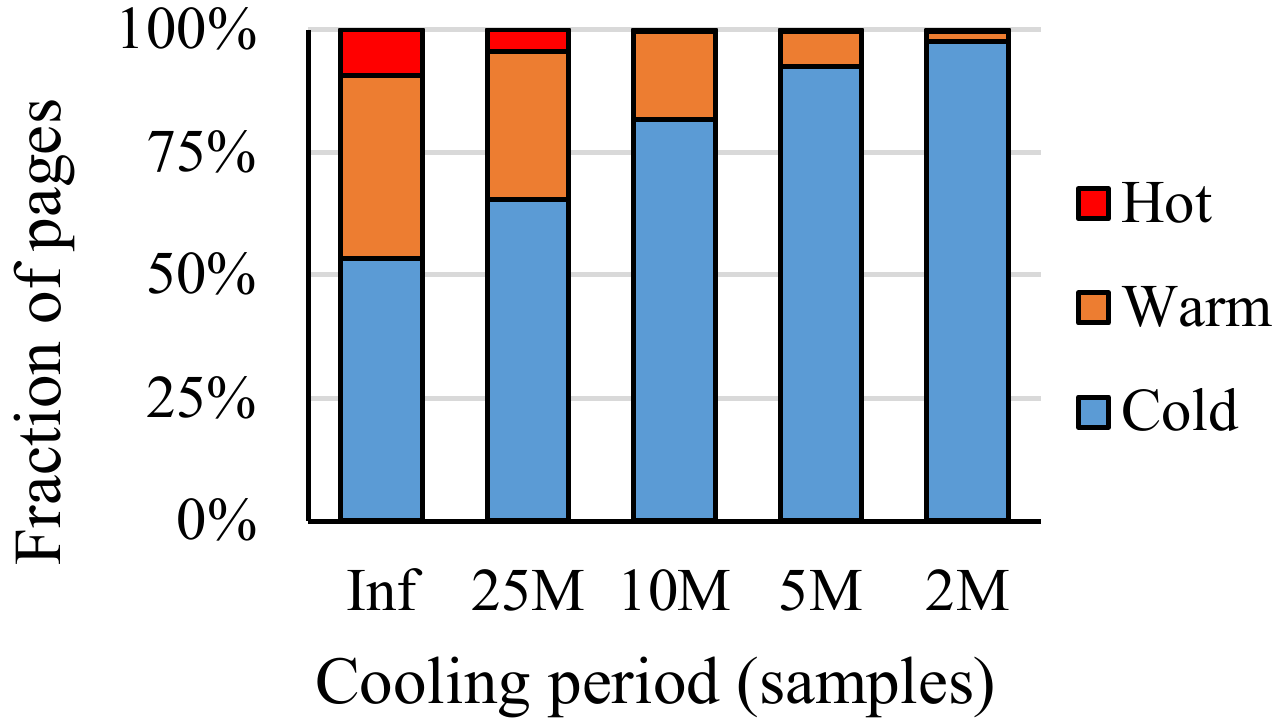}}
  \caption{Effect of cooling period on (a) adaptiveness to hotness changes and (b) hotness classification accuracy. Higher cooling periods adapt to changes more slowly. Lower cooling periods capture hot pages less accurately.}
\label{fig:ema_drawback}
\Description{}
\end{figure}

While frequency-based systems can accurately capture the long-term hotness distribution, they are \textit{suboptimal at quickly adapting to changes in hotness}.
This is because moving average metrics, including EMA, are lagging indicators.
%
%
Intuitively, average scores have ``inertia'' and resist change since they include historical values.
Consider a memory page that was historically hot but recently turned cold. 
Ideally, this page should be demoted quickly to free up space in the fast-tier memory.
The bottom series in \autoref{fig:ema_drawback} (a) represents the access count per minute of such a page. 
The top series represents the EMA score of this page, with cooling performed every 2 minutes, representative of the cooling thresholds used by prior works.
After the 10-minute mark, this page is no longer accessed.
However, since the EMA score is only reduced by half every 2 minutes, it lags behind the access frequency and only drops below 10 after 19 minutes.
This means the tiering system will not be able to identify this page as cold until 9 minutes after it turns cold.
%
%

To adapt faster, one potential solution is to reduce $C$. 
A lower $C$ indeed has a lower lag, as the EMA scores are refreshed more frequently.
However, as demonstrated by prior works \cite{memtis, hemem}, lower values of $C$ also capture the overall hotness distribution less accurately, degrading application performance.  
Intuitively, a lower value of $C$ reduces the number of memory accesses reflected in the overall hotness histogram.
To demonstrate, we measure the hotness distribution of a CacheLib workload under different values of $C$. 
In \autoref{fig:ema_drawback} (b), C=Inf represents the target distribution that should to be captured.
As $C$ decreases, the distribution becomes less accurate because hot and warm pages do not have enough time to accumulate their access counts. 
Therefore, while reducing $C$ improves adaptability, it undermines the tiering system's ability to identify hot data (requirement 1).

\begin{figure}[t]
\centering
\includegraphics[width=0.85\linewidth]{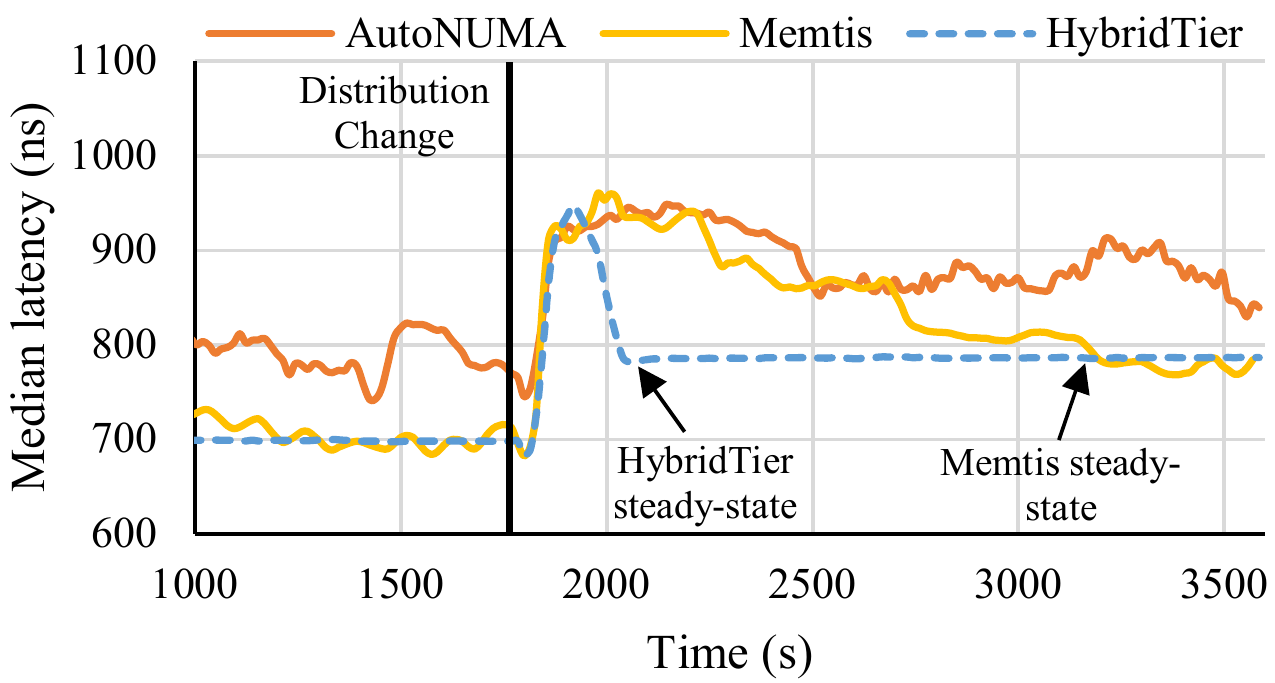}
\caption{Tiering systems adapting to hotness distribution change for CacheLib workload.
  Vertical line indicates when the change in distribution occurs. Lower is better.}
\vspace{-0.5em}
\label{fig:change_dist}
\Description{}
\end{figure}

To evaluate the adaptability of frequency-based tiering systems, we utilize CacheLib, an in-memory cache workload from Meta~\cite{cachelib_paper}.
At the start of the experiment, 100 million cache items are accessed based on a Zipf distribution\footnote{This distribution is provided by Meta as a part of its CacheLib benchmarking framework and reflects Meta's real production environments.}.
We reproduce the varying distribution reported by Meta~\cite{cachelib_paper} by adjusting the access distribution at the 1800-second mark such that 2/3 of previously hot data are no longer hot.
%
\autoref{fig:change_dist} shows that Memtis requires roughly 1400 seconds to adapt to the new distribution, 1000 seconds slower than ideal. 
%
%


\textbf{Recency-based Systems}
On the other hand, systems such as AutoNUMA~\cite{autonuma_huang} and TPP~\cite{tpp} measure data hotness by its access characteristics over the duration of seconds.
For example, AutoNUMA uses the page hint fault time to decide whether a page should be promoted.
AutoNUMA periodically scans the application address space and unmaps 256MB of pages. 
The time elapsed between when an unmapped page is accessed and when it was unmapped is the hint fault latency.
If a page has hint fault latency of less than 1 second, it is promoted, regardless of its historical access statistics. 
The main drawback of recency-based systems is their inability to accurately identify hot pages.
Prior works have shown that migration decisions solely based on recency statistics can be suboptimal \cite{memtis, hemem, multiclock}.
Intuitively, a recently accessed page may or may not be a hot page. 
For example, a cold page with only a single recent access may be misclassified as a hot page by AutoNUMA.
This can be seen by the fact that in \autoref{fig:change_dist}, AutoNUMA continues to have high latency even when the access distribution is no longer changing. 
\autoref{fig:change_dist} also shows that AutoNUMA adapts the slowest to the change in hotness distribution.
This is counter-intuitive at first since recency-based systems do not suffer from the same adaptability drawbacks as frequency-based systems.
The reason behind this is that while AutoNUMA can quickly promote new hot pages, it also incorrectly promotes cold pages at the same time.
In configurations where the fast-tier capacity is limited, these cold pages consume previous fast-tier space and prevent truly hot pages from being promoted. 
From this observation, we argue that accurately capturing hot data (requirement 1) is a prerequisite for high adaptiveness (requirement 2).

\begin{tcolorbox}[]
  \textbf{Observation 1}: Real-world workloads often exhibit skewed but varying hotness distributions. Tiering systems should accurately capture the overall hotness distribution while quickly adapting to changing data hotness. 
\end{tcolorbox}

\subsubsection{Tiering Metadata Overhead} \label{sec:mem_overhead}
Memory tiering systems typically maintain historical access information for each memory page, such as the number of accesses, in order to make future promotion/demotion decisions.
We refer to data structures used to store access information as \textit{tiering metadata}.
We break down the overhead due to tiering metadata into two types: memory overhead and cache overhead.

\textbf{Memory Overhead: }
A common approach used by prior systems is to allocate dedicated metadata for each memory page in the system.
However, a modern large memory server can contain up to billions of 4KB pages.
Storing additional tiering metadata for each page can easily consume gigabytes of memory.
While this overhead might be acceptable on a small scale, it can noticeably impact the cost-effectiveness of tiered memory at the data center scale.
Thus, as CXL memory is expected to further increase the amount of memory per server, tiering systems must rigorously optimize the size of metadata associated with each page~\cite{pagemetadata0, pagemetadata1, pagemetadata2}.

However, the metadata memory overhead of existing works is high.
For example, for every 4KB memory page, Memtis adds 16B of metadata for each Linux struct page.
For a server with 1TB of memory, a 0.39\% memory overhead translates to 3.9GB.
In comparison, the Linux kernel (v6.2) only consumes about 400MB of memory upon boot\footnote{Measured on 1TB server. Includes kernel code, data, and slabs.}. 
To illustrate the potential cost of this overhead, we use a small-scale virtual machine serving as an example. 
The AWS t2.nano instance with 0.5GB memory each at \$4.18/month \cite{aws_t2}.
Assuming a data center with 100,000 such servers, the total memory overhead could have been used to generate \$31.8M per year.
%



\begin{tcolorbox}[]
\textbf{Observation 2}: As the total main memory capacity increases, tiering metadata can lead to prohibitively high memory overhead, resulting in reduced tiering cost-effectiveness. 
\end{tcolorbox}

\textbf{Cache Overhead:}
%
As discussed in \ref{sec:mem_overhead}, tiering metadata on a typical system can consume GBs of memory, larger than the capacity of last-level cache on most systems. 
Frequently updating this metadata can generate cache traffic that interferes with the application.

To efficiently track which memory pages are accessed, prior works have proposed various memory access tracking mechanisms, including utilizing page faults \cite{tpp, autonuma_huang}, page table scanning \cite{multiclock}, and hardware performance counters \cite{memtis, hemem}.
In particular, hardware-counter-based access tracking is promising due to its accuracy and scalability \cite{memtis, hemem}. 
Dedicated event sampling hardware, such as Processor Event-Based Sampling (PEBS) for Intel and Instruction Based Sampling (IBS) for AMD processors, provides a stream of sampled events at a specified frequency. 
Each sampled event contains the exact virtual address accessed by the application and whether it was in local DRAM or CXL memory.

\begin{figure}[t]
  \centering
  \subfloat[Under 4KB pages.]{\includegraphics[width=0.24\textwidth]{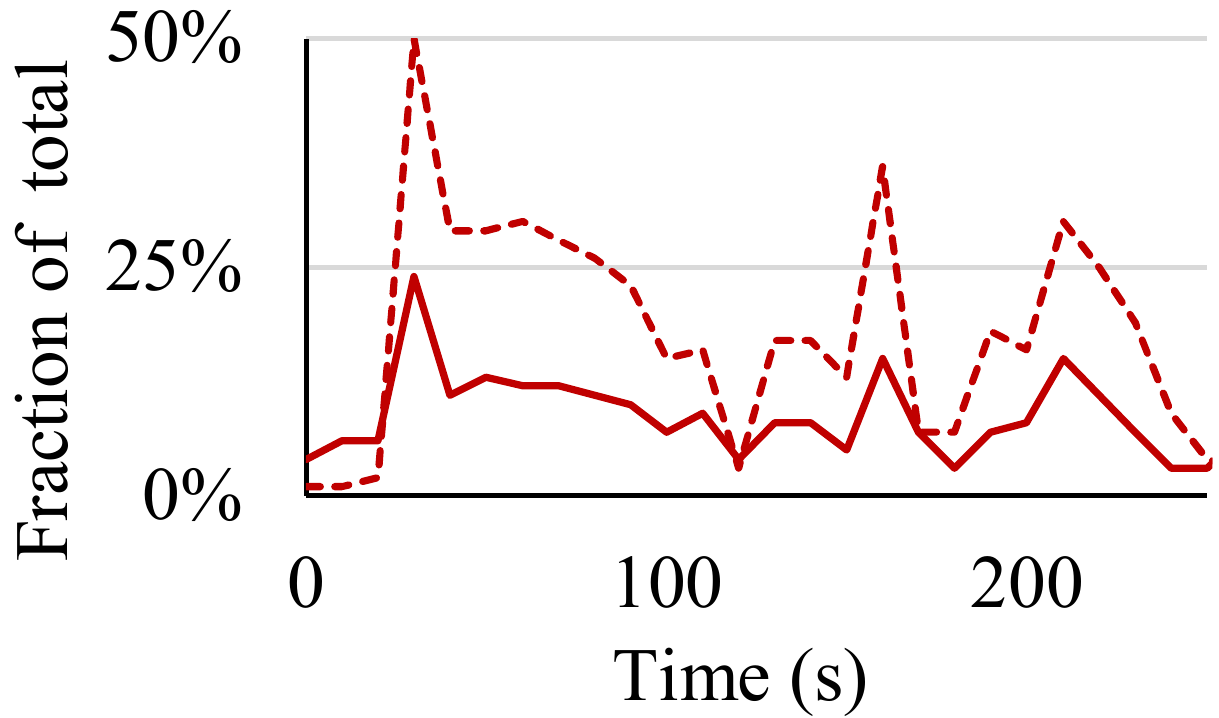}}
\hspace{0.3mm}
  \subfloat[Under huge pages.]{\includegraphics[width=0.23\textwidth]{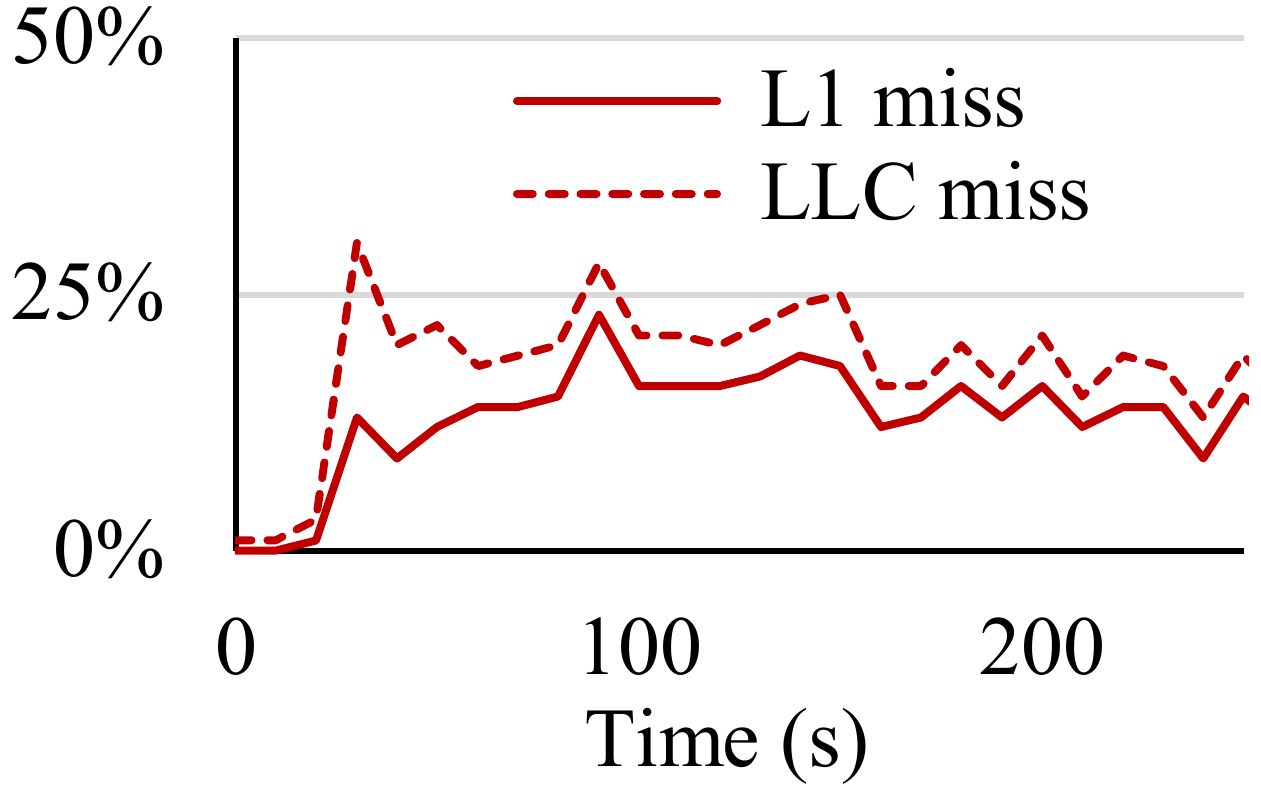}}
  \caption{Cache misses due to Memtis tiering activities as a fraction of the system total when running CacheLib.}
\label{fig:memtis_overhead}
\Description{}
\end{figure}

Despite dedicated sampling hardware, access tracking can have non-negligible caching overhead, which has been overlooked by prior tiering systems. 
%
To measure the cache overhead of Memtis \cite{memtis}, we run the CacheLib workload on a 1:4 configuration. 
We provide the detailed experiment configuration in Section \ref{sec:cache_overhead}.
As Figure \ref{fig:memtis_overhead} shows, tiering activities in Memtis incur a significant number of cache misses.
Memtis on average consumes 9\% and 18\% of total L1 and LLC cache misses for regular pages, and 13\% and 18\% for huge pages.
%
%
Under cache-intensive applications, this large number of cache misses causes cache and memory resource contention, thus degrading performance.
In Section \ref{sec:cache_overhead_solution}, we analyze the main source of cache misses due to tiering.

%
%


\begin{tcolorbox}[]
    \textbf{Observation 3}: Hardware-counter-based memory access tracking can incur non-negligible cache overhead due to frequent metadata updates.
\end{tcolorbox}

%% file: sec/03_overview.tex
\section{\projectname{} Key Ideas} \label{sec:keyideas}

In this section, we summarize the key challenges and how \projectname{} addresses them. 

\subsection{Adapting to Varying Hotness Distributions}

%
Frequency-based tiering can effectively capture the overall hotness distribution, but cannot quickly adjust to changes.
Recency-based tiering can identify new hot pages quickly, but cannot accurately capture the entire hot set, since they do not consider a page's hotness history.
We observe that this tradeoff is the consequence of the fact that \textit{prior systems only tracks one metric for each page}.
For Memtis, this metric is the exponential moving average score, which is a lagging indicator. 
For AutoNUMA, this metric is the hint fault latency, which does not capture long-term access information.


\begin{table}[t]
\small
\setlength{\tabcolsep}{3pt}
    \centering
\caption{HybridTier promotion/demotion policies. A page is considered to have high frequency/momentum if its frequency/momentum is above the corresponding thresholds.}
\label{tab:matrix}
    \begin{tabular}{|l|l|l|}
    \hline
    \diagbox[width=\dimexpr \textwidth/8+2\tabcolsep\relax, height=0.8cm]{Momentum}{Frequency} & High & Low \\
    \hline
    High &  Promote / No Action & Promote / No Action  \\
    \hline
    Low  & Prompt / 2nd Chance & No Action / Demote \\
    \hline
    \end{tabular}
\end{table}


\textbf{\textit{Key Idea: }}
Based on this observation, our key idea is to maintain \textit{two} separate metrics for each page instead of a single metric.
We refer to the two metrics as access ``frequency'' and ``momentum''.
Page access frequency tracks historical access frequency in  the order of minutes to hours.
Access momentum monitors page access intensity in within seconds.
To achieve this, \projectname{} adopts EMA and sets a high EMA cooling period $C$ for frequency counters and a low $C$ for momentum counters.
The key difference between \projectname{} and prior EMA-based tiering systems is that two dedicated EMA counters enable \projectname{} to accurately capture the long-term hotness distribution while \textit{simultaneously} quickly adapts to hotness changes.


Tracking both frequency and momentum enables a flexible migration policy for \projectname{}, shown in \autoref{tab:matrix}. 
%
%
\projectname{} maintains two hotness thresholds: one for frequency and one for momentum.
\projectname{} automatically adjusts the frequency threshold based on the current hotness distribution and fast-tier capacity size, similar to Memtis \cite{memtis}.
The momentum threshold is determined empirically.
\projectname{} promotes pages with high frequency \textit{or} high momentum. 
The heuristic behind this policy is that pages accessed intensely over a short period will likely continue being accessed. 
This enables \projectname{} to quickly promote pages that were cold in the past but recently became hot. We demonstrate the effectiveness of this heuristic in Section~\ref{sec:detail1}.

For demotion, \projectname{} immediately demotes pages with low frequency and low momentum. 
Pages with high momentum but low frequency will not be demoted since they may have been recently promoted. 
Pages with high frequency but low momentum are given a \textit{second chance} to account for pages that are only cold temporarily. 
Instead of immediately demoting such pages, \projectname{} marks and revisits them for a second chance.
%
%
By adopting this tiering policy, \projectname{} adapts to new access distribution the fastest (\autoref{fig:change_dist}) using only 250 seconds.

While maintaining two metrics per page appears simple, naively applying this technique would double the amount of memory consumed by metadata, exacerbating the metadata overhead problem discussed in Section \ref{sec:mem_overhead}.
Next, we discuss our key idea to significantly reduce this overhead.

\subsection{Metadata Memory Overhead}
\label{sec:reducing_num_metadata}


Prior frequency-based tiering systems such as Memtis \cite{memtis} and HeMem \cite{hemem} utilize hash table-like data structures to store page access counts.
We categorize such data structure as \textit{exact data structures}. 
An exact data structure guarantees that a lookup will always return the previous latest value inserted.
For instance, \texttt{hashtable.lookup(key1)} is guaranteed to return \texttt{value1} if the last insertion on \texttt{key1} was \texttt{hashtable.insert(key1, value1)}.
The hash table maintains an exactness guarantee by allocating dedicated memory for each item inserted and by resolving hash conflicts.
However, as discussed in Section \ref{sec:mem_overhead}, when a large number of items are inserted, its memory footprint also becomes large.

\textbf{\textit{Key Idea: }}
%
%
%
In the context of tracking memory accesses, we argue that exactness is not a requirement for achieving high tiering performance. 
Intuitively, even if the access count of a very hot page is off by 1 or 2, it will most likely still be classified as a hot page and no migration decisions will be affected.
Exactness only affects migration decisions in rare cases where the access count of a page is near the hotness threshold. 
We show that such cases are indeed rare and have negligible impact on performance in practice in Section \ref{sec:detail2}.

Following this observation, one of \projectname{}'s key ideas is to utilize \textit{probabilistic} data structures to track page accesses. 
Specifically, \projectname{} adopts counting bloom filters (CBFs) for its frequency tracker and momentum tracker. 
Unlike exact data structures, CBF tracks access counts probabilistically, i.e. "the access count of this page is \textit{probably} $x$ with probability $p$."
Rather than allocating dedicated memory for every page, CBF allocates a fixed-size array of metadata that is shared by all inserted pages. 
\projectname{} utilizes this property to allocate only enough memory to store metadata that is actively used. 
%
%
In addition, \projectname{} only allocates 4 bits per access counter, allowing a maximum count of 15. 
The heuristics behind this approximation is that pages with access count $\geq$ 15 should all be placed in fast-tier memory, thus there is no need to differentiate between them.
We demonstrate that in Section \ref{sec:detail2} this approximation is accurate for all workloads we evaluate.

\begin{algorithm}[t]
\small
\DontPrintSemicolon
    \While{true}{
         \uIf
        {SampleBuffer is not empty} {            
            Sample = SampleBuffer.read() \\
            PageAddr = Sample.addr \\
            Table[PageAddr]->accesses++ \\
        }
    }
\caption{Typical access sampling algorithm used by sample-based tiering systems.
}
\label{algorithm:memtis_sampling}
\end{algorithm}

\subsection{Tiering Cache Overhead} \label{sec:cache_overhead_solution}
To understand why prior hardware-counter-based systems incur high cache overhead, we present the algorithm used by Memtis \cite{memtis} and HeMem \cite{hemem} for access sampling in Algorithm \ref{algorithm:memtis_sampling}. When new access samples are available in the PEBS buffer (line 2), they are collected one by one (line 3). For each sample collected, its page address is extracted (line 4) and the page access count is updated (line 5). For Memtis, this table is the Linux page table\cite{memtis_pagetable}, while HeMem uses a custom hash table \cite{uthash}.
The main source of cache overhead occurs in line 5. For every sample collected, the tiering thread performs a table lookup (\texttt{Table[PageAddr]}). For Memtis, this requires traversing the Linux multi-level page table \cite{memtis_pagetable}, potentially causing multiple cache misses. For HeMem, since its hash table implements, a lookup may result in multiple pointer dereferences to resolve hash collisions. Since the size of this table can easily exceed the LLC cache size (Section \ref{sec:mem_overhead}), frequent metadata accesses result in a large number of cache misses, as shown in Figure \ref{fig:memtis_overhead}.

\textbf{\textit{Key Idea: }}
The 4-bit CBF introduced in the previous section not only reduces memory overhead but also reduces cache overhead for two reasons. First, 4-bit CBF is more compact. \projectname{} assigns at maximum 4$\times$4-bit counters to a memory page, meaning that each 64B cache line can store access counts for at least 32 pages. In contrast, Memtis requires 16B of metadata per page, allowing only 4 metadata per cache line. Second, the CBF is a single-level key-value data structure that intentionally allows hash collisions (details in section \ref{sec:cbf_detail}), therefore reducing the number of pointer dereferences per lookup.

However, the standard counting bloom filter can still cause high cache misses. As we will show in Section \ref{sec:cbf_detail}, a lookup in the standard CBF performs $k$ accesses to retrieve $k$ counters associated with a page. In the worst case, this incurs $k$ cache misses.
To address this, \projectname{} adopts blocked CBF \cite{caffeine_bcbf, bcbf}, an optimization that ensures each CBF lookup will incur exactly one cache access and at most one cache miss.  We describe blocked CBF in more detail in Section \ref{sec:cbf_detail}.



%% file: sec/04a_design.tex
\section{\projectname{}}

In this section, we describe the detailed design of \projectname{}.

\subsection{Workflow Overview}

\begin{figure}[t]
\centering
\includegraphics[width=0.48\textwidth]{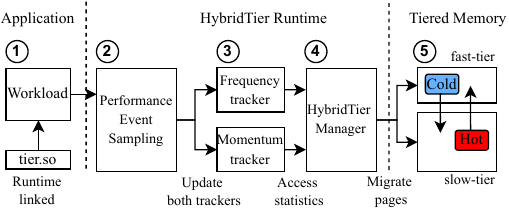}
\caption{An overview of \projectname{}.}
\vspace{-4mm}
\label{fig:highlevel}
\Description{}
\end{figure}

\autoref{fig:highlevel} illustrates the high-level design of \projectname{}. \projectname{} is implemented as a single userspace runtime thread that performs tiering for a workload process.
\circledtext{1} \projectname{} dynamically links the \projectname{} shared library into the target application binary using \texttt{LD\_PRELOAD} dynamic link mechanism. This process is \emph{transparent} to the application and does not require recompiling the target workload.
\circledtext{2} \projectname{} utilizes Intel Processor Event-Based Sampling (PEBS), where each access sample contains the virtual memory address being accessed.
\circledtext{3} \projectname{} stores access statistics using two CBFs, one for each access tracker. 
For each sample collected, \projectname{} updates the access count of the accessed page in both CBFs.  
\circledtext{4} The \projectname{} manager utilizes access statistics from the two trackers to make migration decisions.
%
%
%
%
\circledtext{5} Finally, \projectname{} utilizes system calls to migrate pages between fast and slow-tier memory.

\subsection{Counting Bloom Filter} \label{sec:cbf_detail}
\begin{figure}[t]
\centering
\includegraphics[width=0.48\textwidth]{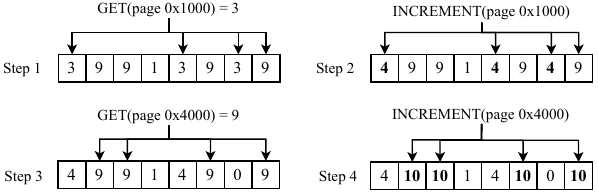}
\caption{Counting bloom filter illustration.}
\vspace{-3mm}
\label{fig:bloom}
\Description{}
\end{figure}

A CBF consists of $k$ hash functions and an array of size $M$. A CBF supports two operations: \texttt{GET} and \texttt{INCREMENT}. 
\texttt{GET} calculates $k$ array indices from $k$ hash functions and returns the minimum counters within the $k$ counters.
\texttt{INCREMENT} calculates $k$ indices from $k$ hash functions and increment the minimum counters. 
\autoref{fig:bloom} shows an example with $k=4$ and $M=8$. At step 1, invoking \texttt{GET} on page 0x1000 will return 3. At step 2, \texttt{INCREMENT} on page 0x1000 will increase the counters at indices 0, 4, and 6 to the value 4. 
Similarly, at step 3, \texttt{INCREMENT} on page 0x4000 will increase the counters at indices 1, 2, 5, and 7 to the value 10, and at step 4 \texttt{GET} on page 0x4000 will return 10. 

The example in \autoref{fig:bloom} shows that the access count of one page may be overwritten by other pages, as the CBF does not resolve hash collisions.
We refer to this error as a ``tracking error''. 
To achieve a balance between tracking error and memory overhead, we compute the size of the CBF $m$ using well-established bloom filter formulas \cite{bloom_calc}:
\begin{gather*} 
r = -k / log(1 - exp(log(p) / k)) \quad m = ceil(n * r) 
\end{gather*}

where $p$ is the probability of tracking error rate, $m$ is the number of counters in the filter. 
In \projectname{}, we empirically set $k=4$, $p=0.001$, and $n$ equal to the number of fast-tier pages. This combination of parameters proved to work well for all our evaluation workloads. 


\textbf{Frequency and Momentum Trackers.} 
Both the frequency and momentum trackers are implemented using CBF.
Since the momentum CBF has a lower cooling period, we observe that the number of pages stored at a given moment is significantly less than that of the frequency CBF. 
This is because the momentum CBF performs cooling frequently, quickly reducing access counts of most pages to 0.
Therefore, we can allocate less memory for the momentum tracker CBF while achieving the desired tracking error.
In practice, \projectname{} allocates 128$\times$ less memory for the momentum CBF than the frequency CBF.

\begin{figure}[t]
\centering
\includegraphics[width=0.42\textwidth]{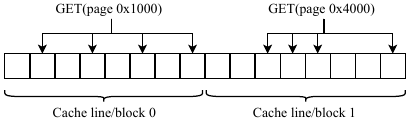}
\caption{Blocked counting bloom filter illustration.}
\vspace{-3mm}
\label{fig:blockcbf}
\Description{}
\end{figure}

\textbf{Blocked CBF.} 
A weakness of the standard CBF illustrated is that the $k$ counters associated with a page lack spatial locality since their memory locations are randomly assigned by the hash functions. In the worst case, a lookup results in $k$ cache misses. Blocked CBF \cite{caffeine_bcbf, bcbf}, illustrated in Figure \ref{fig:blockcbf}, addresses this by enforcing that all $k$ counters of a page are located in the same 64B cache line. A page can be mapped to any but only one cache line. The $k$ counters can be mapped to any counters within the cache line. For illustration, Figure \ref{fig:blockcbf} shows each cache line contains 8 counter slots. In reality, each cache line in a 4-bit CBF contains 128 counter slots. Compared to standard CBF, blocked CBF has a slightly higher false positive rate \cite{caffeine_bcbf, bcbf}. However, in practice, we find the performance benefits to be a favorable tradeoff.

\subsection{Promotion and Demotion} \label{sec:promo_demo_details}

\textbf{Promotion.} 
For each access sample collected, \projectname{} records this access in both CBFs. 
Based on the updated page frequency and momentum, \projectname{} decides whether to promote this page or not.
To reduce system call overhead, \projectname{} processes 100,000 samples as a batch and promotes all hot pages in a batch with a single system call. 

\textbf{Demotion.}
%
When the amount of free memory in the fast-tier is below \texttt{PROMO\_WMARK} watermark, \projectname{} demotes until the amount of free memory in the fast-tier is greater than \texttt{DEMOTE\_WMARK}. 
\projectname{} identifies cold pages in fast-tier by linearly scanning the application virtual address space utilizing \texttt{/proc/PID/maps} and \texttt{/proc/PID/pagemaps}.
When a page is marked for second-chance, its current access frequency is saved. 
\projectname{} later revisits previously marked pages and compares their current access frequency count against the previously stored count. 
If a marked page was not accessed after the revisit time, \projectname{} considers this page to be no longer hot and demotes it.
We empirically set the revisit time to 1 minute to achieve a balance between demotion accuracy and runtime overhead.

\subsection{Huge Page Support}
\projectname{} supports 2MB huge pages through Linux Transparent Huge Pages (THP).
When enabled, \projectname{} tracks access frequency/momentum and performs migrations at the huge page granularity.
\projectname{} increase each CBF counter to 16-bit CBFs to accommodate higher access counts for huge pages.
At the same time, the number of elements in each CBF is also reduced as the total number of pages in the system is reduced by 512$\times$.
Therefore, \projectname{}'s metadata memory consumption in huge page mode is 128$\times$ lower than in regular page mode.

%% file: sec/04b_implementation.tex
\subsection{Implementation Details} \label{sec:impl}
\projectname{} is a userspace runtime thread.
We implement \projectname{} using 1,577 lines of code in C++.
As a userspace runtime, \projectname{} requires no workload recompilation and kernel modifications. 
\projectname{} relies on the operating system and hardware support for (1) memory access sampling, (2) memory movement between tiers, and (3) the ability to scan pages in a process address space. 
These requirements are widely available in modern systems, such as Instruction-Based Sampling (IBS) in AMD processors \cite{ibs} and NUMA migrations in Windows systems \cite{windows_perf}. 

%% file: sec/05_methodology.tex
\section{Methodology}

\subsection{CXL Emulation} 

While recent studies have revealed performance characteristics of real CXL-memory devices (CXL 1.1 specification)~\cite{uiuc_cxl}, many of these CXL devices are not commercially available on the market.
Similar to recent works~\cite{tpp, pond, memtis}, we use a remote NUMA node on a two-socket system to emulate CXL.
The emulated CXL memory has idle latency of 124ns and bandwidth of 34 GB/s, similar to reported in a recent work~\cite{uiuc_cxl}.
Each socket has a 16-core Intel Xeon 4314 processor and 512GB of DDR4 memory. 
The application runs only on local NUMA node CPUs.

\begin{table}[t]
\small
    \centering
    \caption{Workloads for evaluation.}
    \setlength\tabcolsep{3pt}
    \aboverulesep = 0.2mm \belowrulesep = 0.2mm 
    \begin{tabular}{l l l} 
     \toprule
     Application & Input  & Footprint \\
     \midrule
     Content-delivery network & \multirow{2}{*}{CacheLib generator~\cite{cachelib_git}} & \multirow{2}{*}{267GB}\\
     
     Social-graph & & \\
     \midrule
     Breadth-first search (BFS) & Kronecker graph~\cite{gap} & \multirow{3}{*}{335GB} \\
     Connected components (CC) & Uniform random graph &  \\
     Page Rank (PR) &  &  \\
     
     \midrule
     \multirow{2}{*}{SPEC CPU 2017} &  603.bwaves~\cite{bwaves}  & \multirow{2}{*}{150GB}\\
      &  654.roms~\cite{roms}  &  \\
     \midrule

     Silo~\cite{silo} &  YCSB-C~\cite{ycsb}  & 208GB \\
     \midrule
     
     XGBoost~\cite{xgboost_git} &  Criteo Click Logs~\cite{criteo_link}  & 248GB \\
     \bottomrule
    \end{tabular}
    \label{table:workloads}
    \vspace{-3mm}
\end{table}

\subsection{Baselines and \projectname{} Configurations}
We compare \projectname{} against AutoNUMA~\cite{autonuma_huang}, TPP~\cite{tpp}, Memtis~\cite{memtis}, ARC~\cite{arc}, and TwoQ~\cite{twoq}. 
For AutoNUMA, \projectname{}, we use Linux kernel v6.2.
%
%
For AutoNUMA, we enable its multi-generational LRU (MGLRU) based demotion due to its better performance than regular LRU. 

We implement two additional tiering systems based on traditional caching algorithms: ARC \cite{arc} and TwoQ \cite{twoq}. ARC \cite{arc} is a self-tuning caching policy that maintains two LRU lists to estimate item recency and frequency. TwoQ \cite{twoq} is an extension of LRU that utilizes two queues to differentiate between items accessed only once vs. multiple times. Since ARC and TwoQ assume that the cache is initially empty, we initially allocate new memory pages on slow-tier memory for these two baselines. 
We omit comparisons against HeMem~\cite{hemem} and Tiering-0.8 \cite{tiering0.8} as Memtis already performs detailed comparisons against it.
We do not perform end-to-end evaluations on MTM~\cite{mtm} as its source code is not publicly available at the time of writing.
%

\subsection{Workloads} 
We evaluate \projectname{} on workloads in Table \ref{table:workloads}.
All experiments use 16 threads mapped to 16 physical cores.
%
%
CacheLib is an in-memory cache used by Meta~\cite{cachelib_paper, cachebench}.
We evaluate two workloads: content delivery networks (CDN) and social graphs.
Each workload is characterized by a custom popularity distribution, size distribution, and operation composition that are representative of production workloads.
GAP is a collection of standard graph processing kernel implementations \cite{gap}.
We generate two graphs: Kronecker graph and uniform random graph~\cite{gap}, each with 2 billion nodes and 8 billion edges. 
The uniform random graph represents the worst case in terms of locality, where every vertex has an equal probability of being a neighbor of every other vertex.
We evaluate three kernels: breath-first search (BFS), connected components (CC), and page rank (PR).
SPEC CPU 2017 is an industry-standard CPU intensive benchmark suite \cite{speccpu}. 
We select 603.bwaves \cite{bwaves} and 654.roms \cite{roms} as they have the largest memory footprints.
We follow the official SPEC CPU 2017 guidelines \cite{bwaves, roms} to scale up their input sizes to achieve ~150GB resident set size.
Silo \cite{silo} is an in-memory database engine. Similar to Memtis \cite{memtis}, we use YCSB-C input workload to stress the database engine. 
XGBoost is a widely used gradient-boosting library implemented using C++ commonly executed on CPU systems~\cite{xgboost_cpu1, xgboost_cpu2}.
We evaluate XGBoost training using the Criteo Click Logs dataset~\cite{criteo_link}.

%% file: sec/06_evaluation.tex
\section{Evaluation} \label{sec:eval}
In this section, we evaluate \projectname{} by performing end-to-end performance comparisons on regular 4KB pages (Section~\ref{sec:main_result_4kb}) and huge pages (\ref{sec:huge_perf}). Then, we perform detailed comparisons against Memtis (\ref{sec:comp_memtis}) and conduct experiments to understand \projectname{}'s performance (\ref{sec:understand_perf}).

\subsection{Regular Page Performance} \label{sec:main_result_4kb}

\begin{figure}[t]
  \centering
  \includegraphics[width=0.9\linewidth]{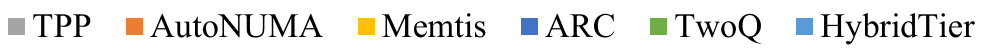}
  \vspace{-3mm}
      \hspace{0.5mm}

  \subfloat[CacheLib CDN workload.]{
  \includegraphics[width=0.45\linewidth]{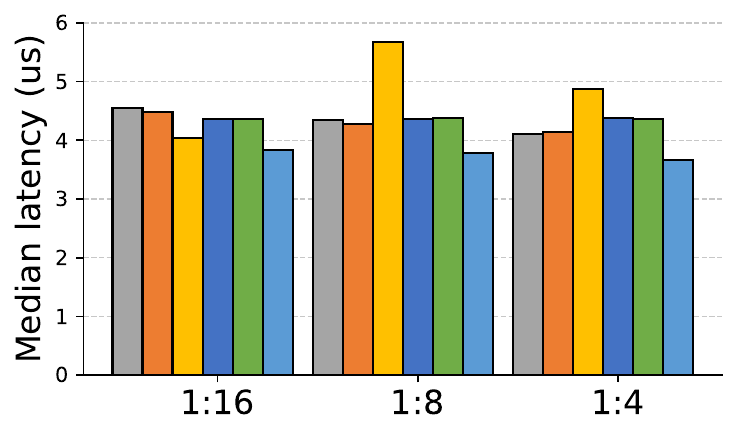}
  \hfill
  \includegraphics[width=0.45\linewidth]{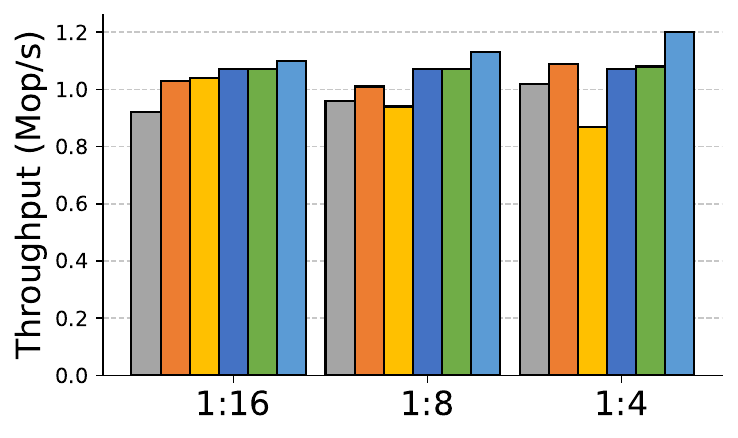}
  }
  \hspace{-0.5mm}
  \vspace{-3mm}
  
   \subfloat[CacheLib Social-graph workload.]{
   \includegraphics[width=0.45\linewidth]{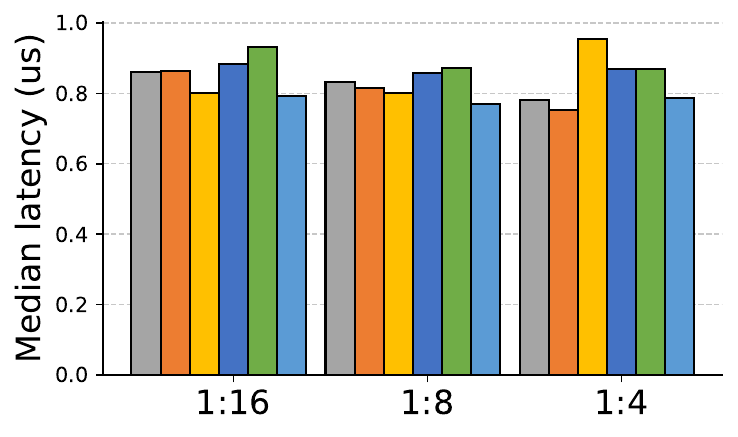}
   \hfill
   \includegraphics[width=0.45\linewidth]{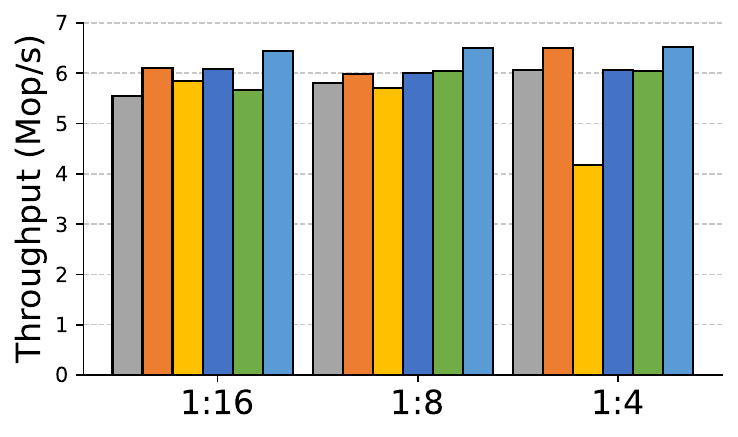}
   }
  \caption{Performance evaluation for CacheLib workloads. Lower is better for latency. Higher is better for throughput.}
\label{fig:cachelib_main}
\Description{}
\end{figure}

\begin{figure}[t]
  \centering
  \includegraphics[width=0.9\linewidth]{results/main_legend.pdf}
    \hspace{0.5mm}
  \vspace{-3mm}

  \subfloat[BFS Kronecker]{\includegraphics[width=0.45\linewidth]{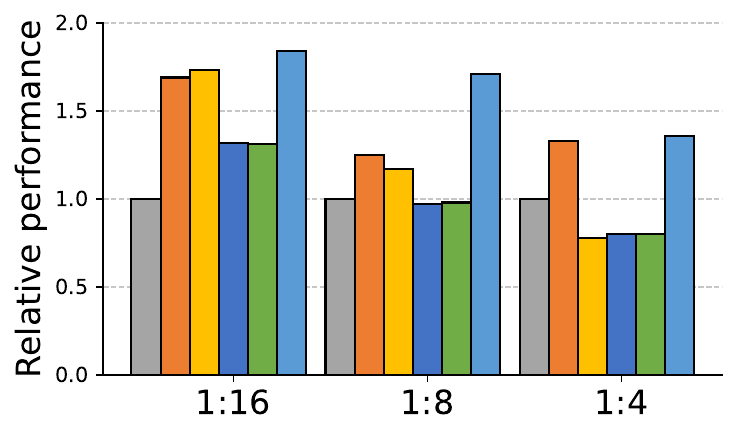}}
   \hfill
  \subfloat[BFS uniform random]{\includegraphics[width=0.45\linewidth]{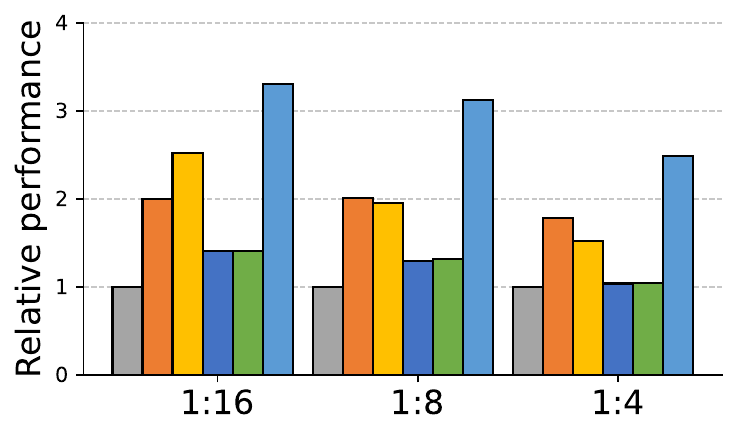}}
  \vspace{-3mm}

  \subfloat[CC Kronecker]{\includegraphics[width=0.45\linewidth]{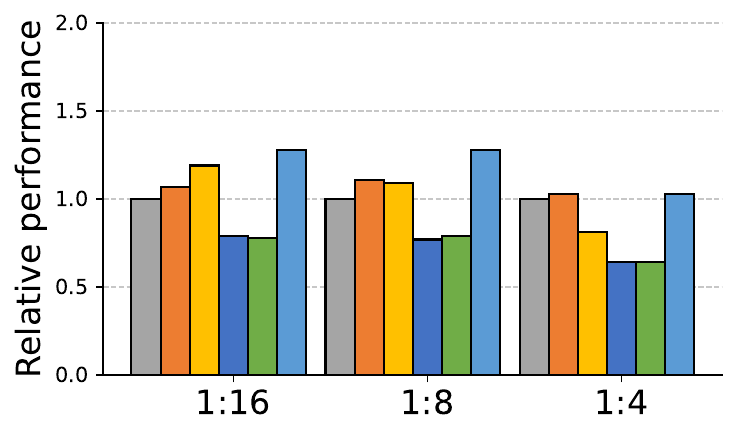}}
    \hfill
  \subfloat[CC uniform random]{\includegraphics[width=0.45\linewidth]{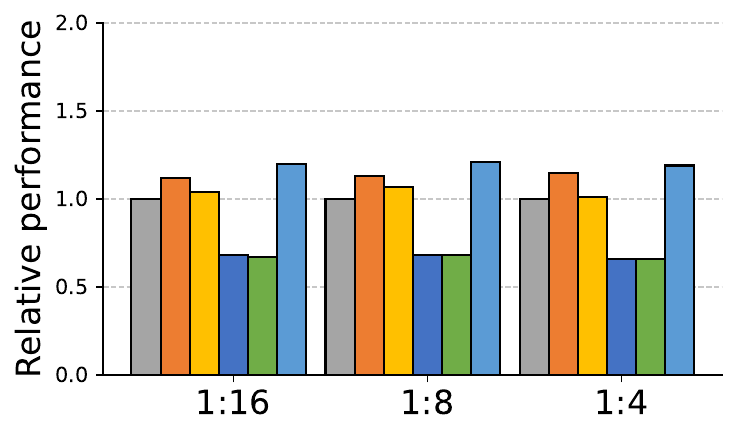}}
  \vspace{-3mm}

   \subfloat[PR Kronecker]{\includegraphics[width=0.45\linewidth]{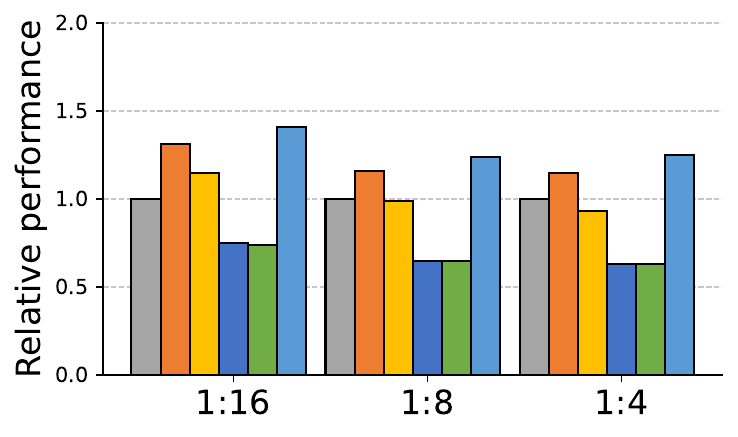}}
    \hfill
    \subfloat[PR uniform random]{\includegraphics[width=0.45\linewidth]{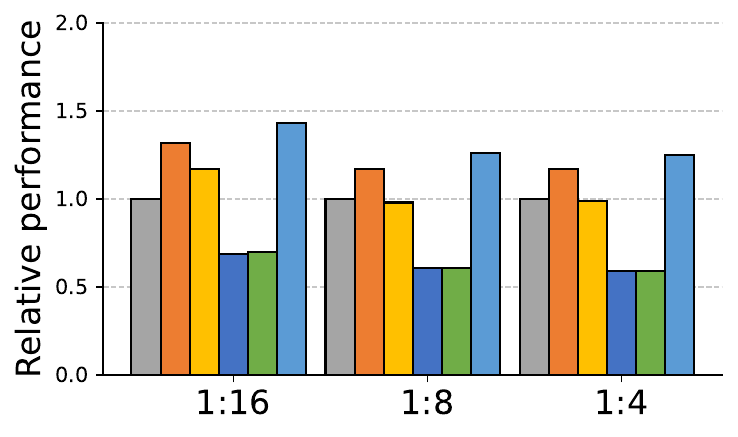}}
    \vspace{-3mm}

   \subfloat[SPEC CPU 603.bwaves]{\includegraphics[width=0.45\linewidth]{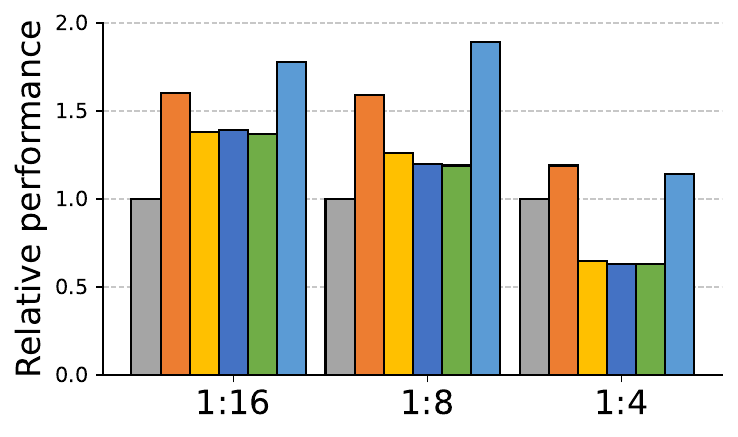}}
    \hfill
    \subfloat[SPEC CPU 654.roms]{\includegraphics[width=0.45\linewidth]{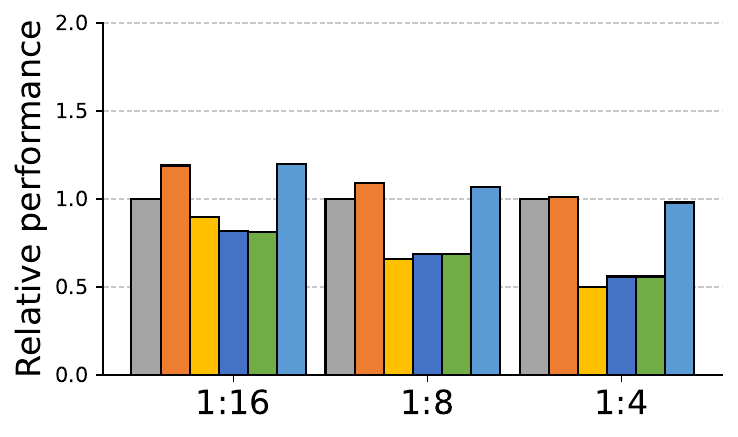}}
    \vspace{-3mm}

   \subfloat[Silo]{\includegraphics[width=0.45\linewidth]{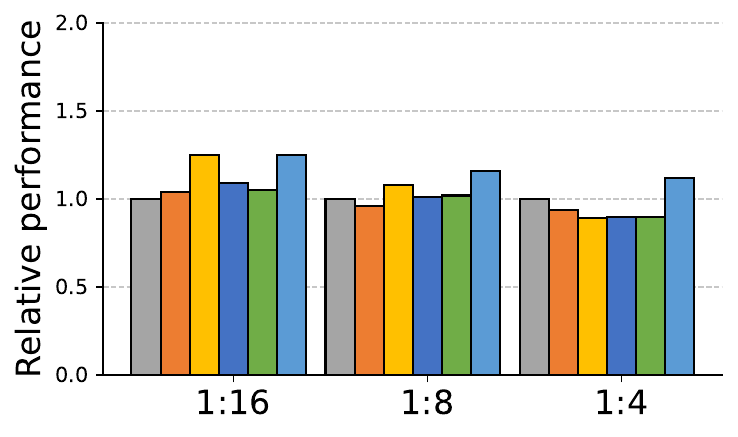}}
    \hfill
    \subfloat[XGBoost]{\includegraphics[width=0.45\linewidth]{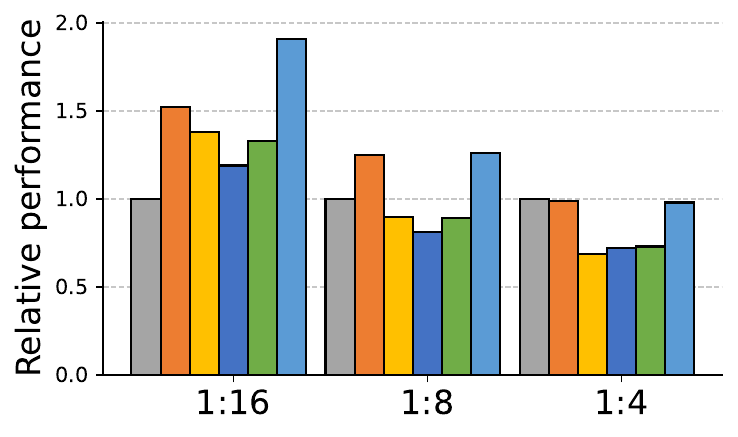}}
    \vspace{-3mm}
    
  \caption{Performance comparison of \projectname{}. All performance normalized against TPP. Higher is better. }
\label{fig:gap_main}
\Description{}
\end{figure}

\autoref{fig:cachelib_main} and \autoref{fig:gap_main} show the performance comparison of \projectname{}. 
The x-axis indicates the ratio between fast and slow-tier memory capacity, where the slow-tier capacity is fixed at 512GB.
On average (geomean), \projectname{} outperforms TPP, AutoNUMA, Memtis, ARC, and TwoQ by 32\%, 11\%, 29\%, 50\%, and 40\% respectively.

\textbf{CacheLib.}
\autoref{fig:cachelib_main} shows the cache access median latency and throughput of \projectname{} and prior works.
We make two observations.
1) Under the same fast:slow memory ratios, \projectname{} performs the best in all but two experiments.
On average, \projectname{} outperforms TPP, AutoNUMA, Memtis, ARC, and TwoQ by 10\%, 9\%, 18\%, 14\%, and 15\% respectively in terms of median latency. \projectname{} also improves throughput by 15\%, 7\%, 23\%, 7\%, and 8\% respectively.
2) \projectname{} often requires 2$\times$ less fast-tier memory to achieve the same level of performance as the second best performing system.
On the CDN workload, \projectname{} with 1:16 configuration outperforms all other systems with 1:8 configuration. 

Compared to the frequency-based Memtis, \projectname{}'s speedups mainly come from its adaptability and low cache overhead.
Compared to recency-based AutoNUMA and TPP, \projectname{} can identify hot pages more accurately.
Surprisingly, Memtis often performs worse with a higher fast:slow ratio.
%
%
We profile Memtis and observe that under larger fast-tier memory, Memtis performs additional background activities that result in higher runtime overhead.

\textbf{GAP.} \autoref{fig:gap_main} (a) to (f) show the relative performance of \projectname{} compared to baselines. 
On average, \projectname{} outperforms TPP, AutoNUMA, Memtis, ARC, and TwoQ by 51\%, 16\%, 29\%, 88\%, and 88\% respectively.
Out of the GAP workloads, \projectname{} shows the largest speedup for BFS, outperforming the second best system by 33\% on average for both input graphs. 
The reason behind this is that BFS is a "single-source" kernel, where a different source vertex is selected for each iteration \cite{gap}.
In contrast, CC and PR are "whole-graph" kernels, where the entire graph is processed the same way every trial.
As a result, the BFS kernel experiences different hotness distributions for different source vertices.
\projectname{}'s adaptive tiering policy can quickly adjust to this change in hotness distribution. 

In terms of absolute runtime, all tiering systems perform worse under uniform random input graph than under Kronecker graph.
This is expected since uniform random graph represents the worst case in terms of locality.
\projectname{}'s speedup over the second-best system grows from 15\% on the Kronecker graph to 53\% on the uniform random graph for BFS. 
This occurs since uniform random graph amplifies variations in node hotness. 
A more uniform graph is more likely to produce diverse hot sets, whereas a more concentrated graph, such as the Kronecker graph, tends to maintain a more consistent hot set of nodes.

In general, ARC and TwoQ show similar performances. 
This is expected, since both ARC and TwoQ use multiple LRU queues to estimate item recency and frequency. 
While in theory, TwoQ can perform worse since it has two parameters that need to be tuned (Kin and Kout), we found that the default values provided by the original paper \cite{twoq} worked well: $K_{in} = maxSize / 4$ and $K_{out} = maxSize / 2$.
Compared to other tiering systems, ARC and TwoQ generally perform worse.
We profile ARC and TwoQ's page migrations and observe that this is mainly because of their lenient promotion policies.
Upon a cold miss (the first time a page is sampled), both systems directly promote the missed page.
We observe that this promotion policy is often too aggressive and can mistakenly promote cold pages.
Except for Memtis, as fast-tier capacity increases, the performance gap between \projectname{} and other baselines reduces, since the penalty for mispromotion is low with abundant fast-tier memory.
Similar to CacheLib, we observe Memtis performance drops at higher fast:slow configurations.

\textbf{SPEC CPU, Silo, and XGBoost.}
On average, \projectname{} outperforms the second best system by 3\%, 20\%, and  8\% for SPEC CPU, Silo, and XGBoost respectively. 
While AutoNUMA has the second best performance on SPEC CPU and XGBoost, it performs worse than Memtis on Silo.
Silo uses YCSB, which uses a workload generator that assumes no changes in hotness, as each key remains equally hot throughout the benchmark.
This is advantageous for the hotness histogram adopted by Memtis, as discussed in Section \ref{sec:capture_hot}.

\begin{figure}[t]
  \centering
\includegraphics[width=1\linewidth]{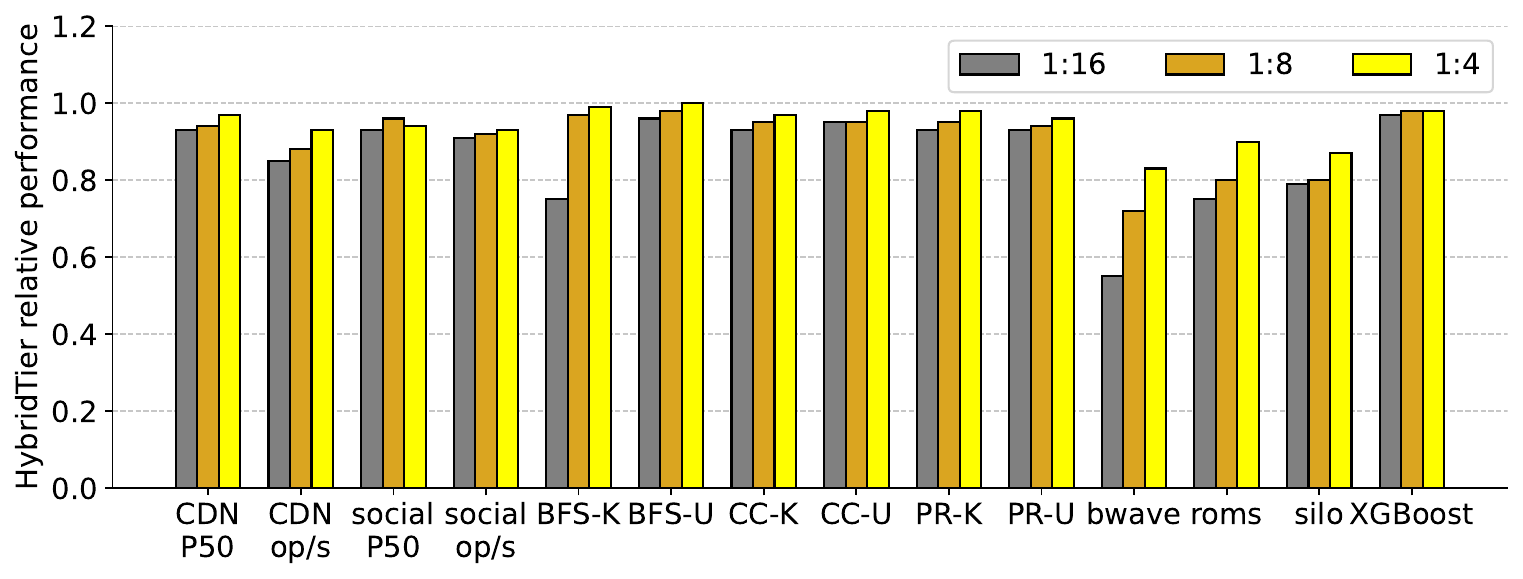}
  \vspace{-5mm}
  \caption{\projectname{} performance normalized against baseline using all fast-tier memory.}
\label{fig:all_local}
\Description{}
\end{figure}

\textbf{Comparison against all fast-tier. }
\autoref{fig:all_local} shows the performance 
of \projectname{} normalized against baseline where only the fast-tier memory is used. 
This represents the performance upper bound of memory tiering systems. 
Under 1:16, 1:8, and 1:4 memory configurations, \projectname{} is on average 14\%, 9\%, and 6\% slower than all fast-tier.
%

\begin{figure}[t]
  \centering
\includegraphics[width=1\linewidth]{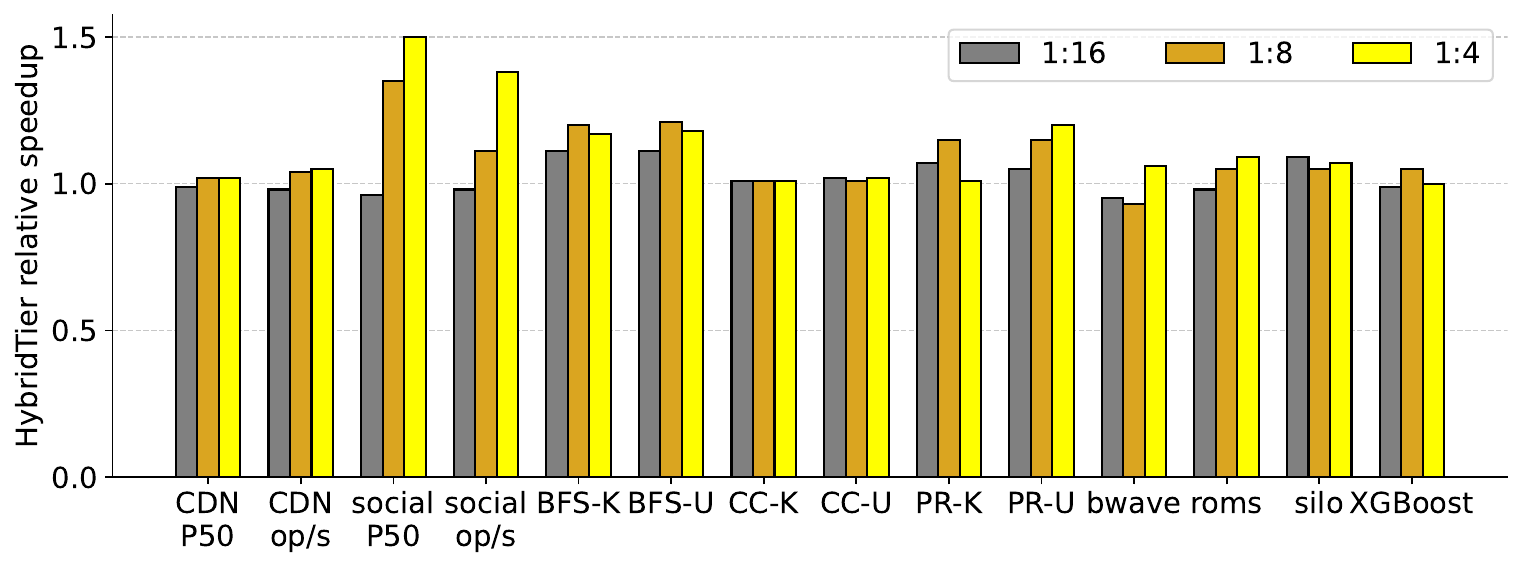}
  \vspace{-5mm}
  \caption{\projectname{} huge page performance normalized against Memtis. Higher is better for \projectname{}.}
\label{fig:main_huge}
\Description{}
\end{figure}

\subsection{Huge Page Performance} \label{sec:huge_perf}
To evaluate \projectname{}'s huge pages performance, we compare against Memtis \cite{memtis} on all workloads.
\autoref{fig:main_huge} shows that on average, \projectname{} outperforms Memtis by 9\% and 11\% for 1:8 and 1:4 configurations while performing on par for 1:16.
\projectname{} shows the most performance improvements over Memtis on CacheLib social-graph, BFS, and PR.

\subsection{Detailed Comparison Against Memtis} \label{sec:comp_memtis}
In this section, we compare \projectname{} against Memtis, the state-of-the-art frequency-based tiering system that also uses PEBS hardware sampling. We compare in terms of adaptiveness to dynamic distributions (Section~\ref{sec:adapt}), metadata memory overhead (\ref{sec:metadata_overhead}), and tiering cache overhead (\ref{sec:cache_overhead}).

\begin{table}[t]
    \small 
    \centering
    \caption{Minutes required to adapt to new access distribution (reach within 1\% of the steady-state median latency). 
    } 
    \vspace{-3mm}
    \setlength\tabcolsep{6pt}
    \aboverulesep = 0.2mm \belowrulesep = 0.2mm 
    \addtolength{\tabcolsep}{-0.25em}
    \begin{tabular}{c  |c  c  c | c c c } 
     \toprule
     & \multicolumn{3}{c|}{CDN} & \multicolumn{3}{c}{Social-graph}  \\
      \midrule
      &  1:16 & 1:8 & 1:4 & 1:16 & 1:8  & 1:4 \\
     \midrule
       Memtis  &>60 & 42.6 & >60  & 34.2 & >60 &  29.6 \\  
     \projectname{}  &	25.6 &	25.2 & 23.4 & 9.6 &	10.1 & 8.9 \\  
     Relative Reduction   &	2.3$\times$ &	1.7$\times$ & 2.6$\times$ &	3.6$\times$ &	5.9$\times$ & 3.3$\times$ \\  
     \bottomrule
    \end{tabular}
    \label{table:adapt}
\end{table}

\subsubsection{Adaptiveness to Dynamic Distributions} \label{sec:adapt}
Table \ref{table:adapt} shows the amount of time required for \projectname{} and Memtis to adapt to a new hotness distribution on two CacheLib workloads.
We measure how long it takes for each tiering system to reach within 1\% of the steady-state median latency. 
On average, \projectname{} requires 3.2$\times$ less time to adapt. 
\projectname{}'s tiering policy considers both long-term access frequency and short-term access momentum, which enables it to quickly capture pages that recently turned from cold to hot and vice versa. 
On the other hand, Memtis must wait for its cooling mechanism to reduce the page's access count, resulting in a long delay before the page can be demoted. 

\subsubsection{Metadata Memory Overhead} \label{sec:metadata_overhead}
\autoref{table:memoverhead} shows the relative amount of metadata incurred by \projectname{} compared to Memtis.
On average, \projectname{} incurs 4.6$\times$ less metadata overhead than Memtis. Since \projectname{}'s metadata size scales with the size of fast-tier memory, it achieves larger memory savings at lower fast-tier sizes. On the other hand, Memtis metadata overhead scales with the total memory capacity and thus remains constant in our setup.

\begin{table}[t]
    \small 
    \centering
    \caption{Size of metadata relative to total memory capacity.  } 
    \vspace{-3mm}
    \setlength\tabcolsep{8pt}
    \aboverulesep = 0.2mm \belowrulesep = 0.2mm 
    \addtolength{\tabcolsep}{-0.25em}
    \begin{tabular}{c c  c  c} 
     \toprule
       & 1:16 & 1:8 & 1:4   \\
     \midrule
     Memtis  &  0.39\% & 0.39\%	& 0.39\% \\  
     \projectname{} 	& 0.050\% &	0.097\% &  0.192\% \\  
     Relative Reduction  & 7.8$\times$ & 4.0$\times$ & 2.0$\times$   \\  
     \bottomrule
    \end{tabular}
    \label{table:memoverhead}
\end{table}

\subsubsection{Cache Overhead} \label{sec:cache_overhead}
\begin{figure}[t]
  \centering
  \subfloat[Under 4KB page.]{\includegraphics[width=0.225\textwidth]{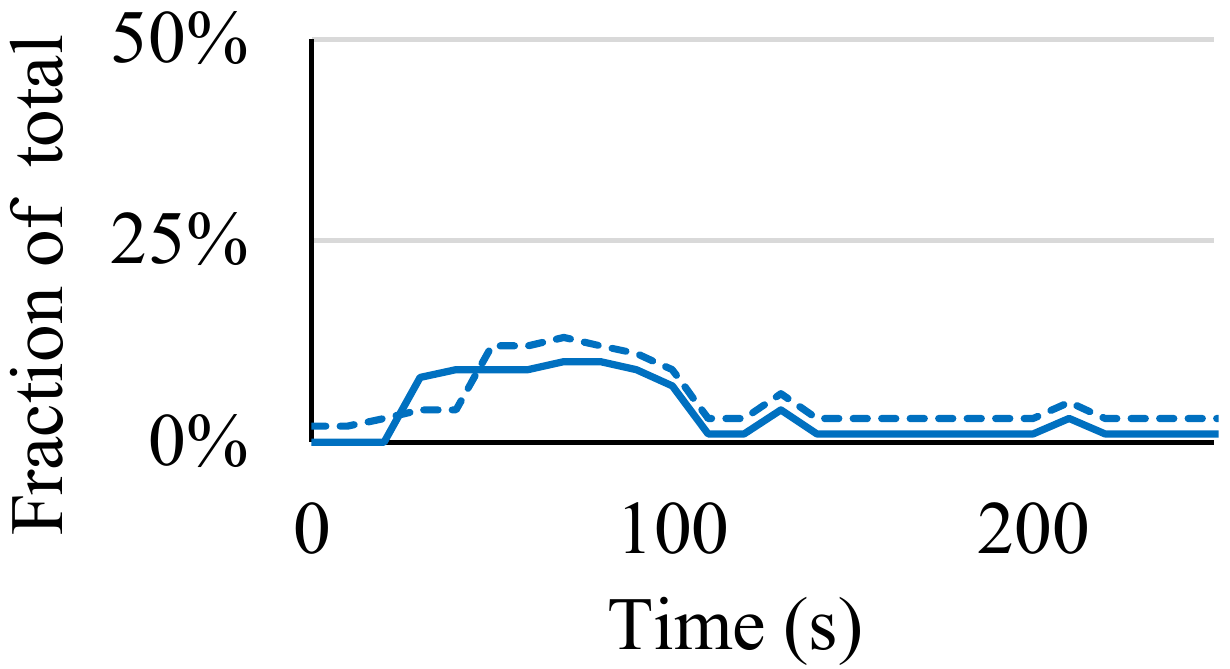}}
  \hspace{0.3mm}
  \subfloat[Under huge page.]{\includegraphics[width=0.228\textwidth]{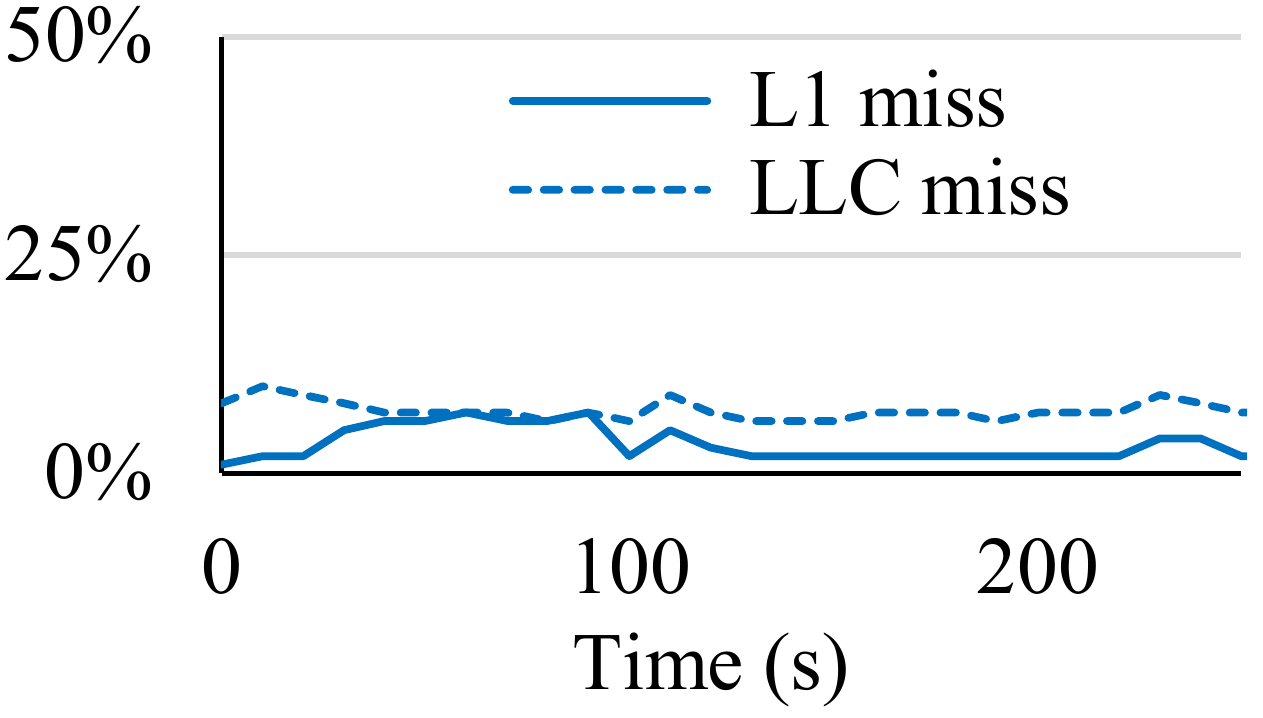}}
  \caption{Cache misses due to \projectname{} 
 tiering activities as a fraction of the system total. }
\label{fig:cache_overhead_result}
\Description{}
\end{figure}

\begin{figure}[t]
  \centering
    \includegraphics[width=0.4\textwidth]{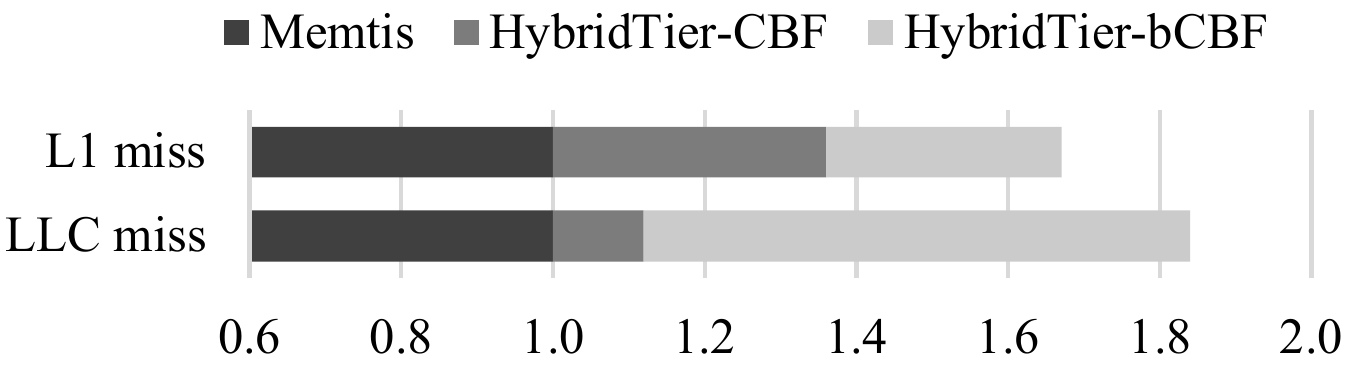}
  \hspace{0.5mm}
  \vspace{-4mm}
  \caption{HybridTier cache miss reduction breakdown.
  }
\label{fig:cache_overhead_breakdown}
\Description{}
\end{figure}

To evaluate tiering cache overhead,
we use \texttt{perf} to separately record the number of cache accesses issued by the application and by tiering threads. 
Figure \ref{fig:cache_overhead_result} shows that on average, \projectname{} generates 5\% and 4\% of total cache misses under regular and huge pages respectively. 
Overall, \projectname{} reduces the total number of L1 and LLC cache misses by 1.7$\times$ and 1.8$\times$ when using regular pages, and 3.2$\times$ and 3.5$\times$ under huge pages.
Figure \ref{fig:cache_overhead_breakdown} breaks down the impact of adopting 4-bit CBF and blocked-CBF under regular pages. 
Using standard CBF (\projectname{}-CBF) moderately reduces cache misses by $12-36\%$ and applying blocked CBF (\projectname{}-bCBF) reduces misses by another $31-72\%$. We conclude that both optimizations are effective while blocked CBF provides a larger reduction.

\subsection{Understanding \projectname{} Performance} \label{sec:understand_perf}
In this section, we analyze the impact of \projectname{}'s key ideas.
Specifically, we discuss 1) how frequency-momentum tracking affects performance (\ref{sec:detail1}) 2) how counting bloom filter impacts tiering accuracy (\ref{sec:detail2}).

\subsubsection{Frequency Momentum Tracking} \label{sec:detail1}
\begin{figure}[t]
\centering
\includegraphics[width=.7\linewidth]{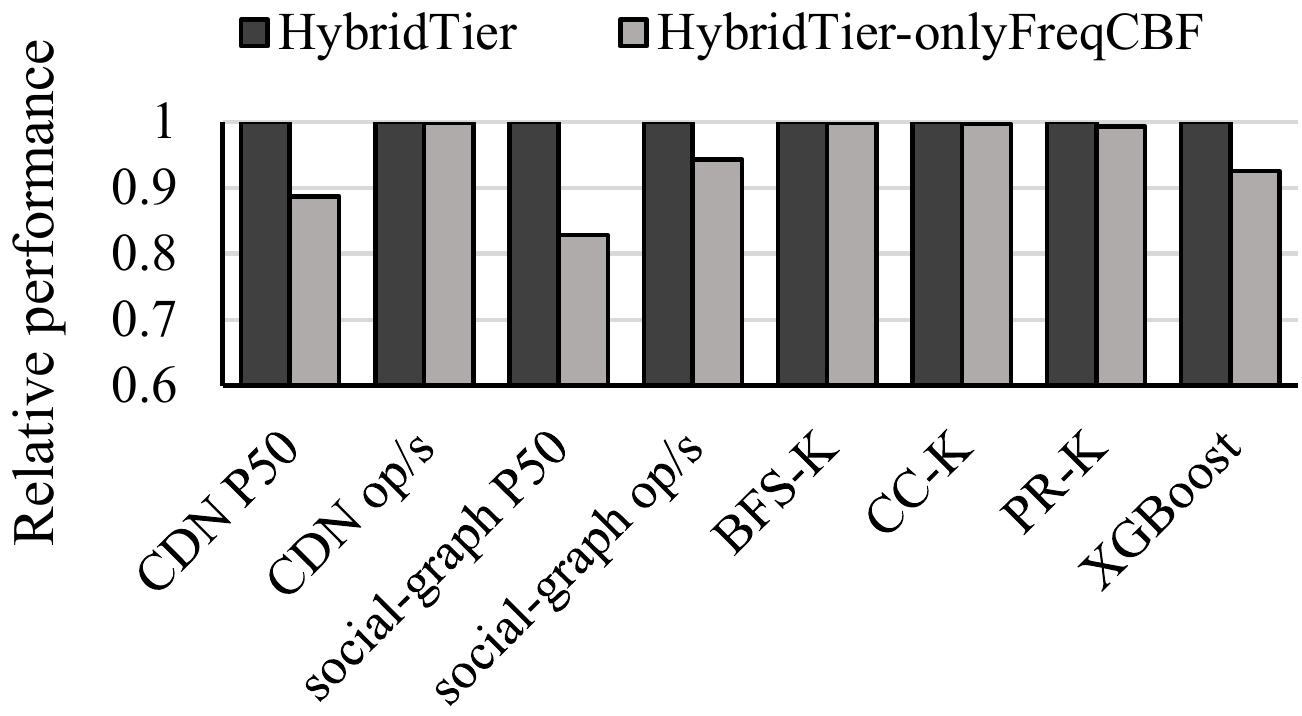}
\caption{Performance of \projectname{} if only the frequency tracker is used (1:8 configuration).}
\vspace{-5mm}
\label{fig:onlyfreq}
\Description{}
\end{figure}

\autoref{fig:onlyfreq} shows that tracking both frequency and recency is most effective for CacheLib and XGBoost, improving their performance by 8.5\% on average.
For BFS, CC, and PR, the performance is similar.
This can be attributed to the fact that these three workloads have small hot sets that can easily fit in fast-tier memory.
\autoref{fig:freq_dist} shows the cumulative access distribution for all workloads we evaluate.
For GAP workloads under Kronecker graph, 94\% of allocated pages have 0 access frequency, meaning that only 6\%, or 20GB of pages are considered warm or hot. 
Since the fast-tier capacity is 64GB, \projectname{} can achieve good performance without the momentum tracker. 

\subsubsection{Counting Bloom Filter} \label{sec:detail2}
\begin{figure}[t]
  \centering
  \subfloat[CacheLib, SPEC CPU, Silo, and XGBoost.]{\includegraphics[width=0.35\textwidth]{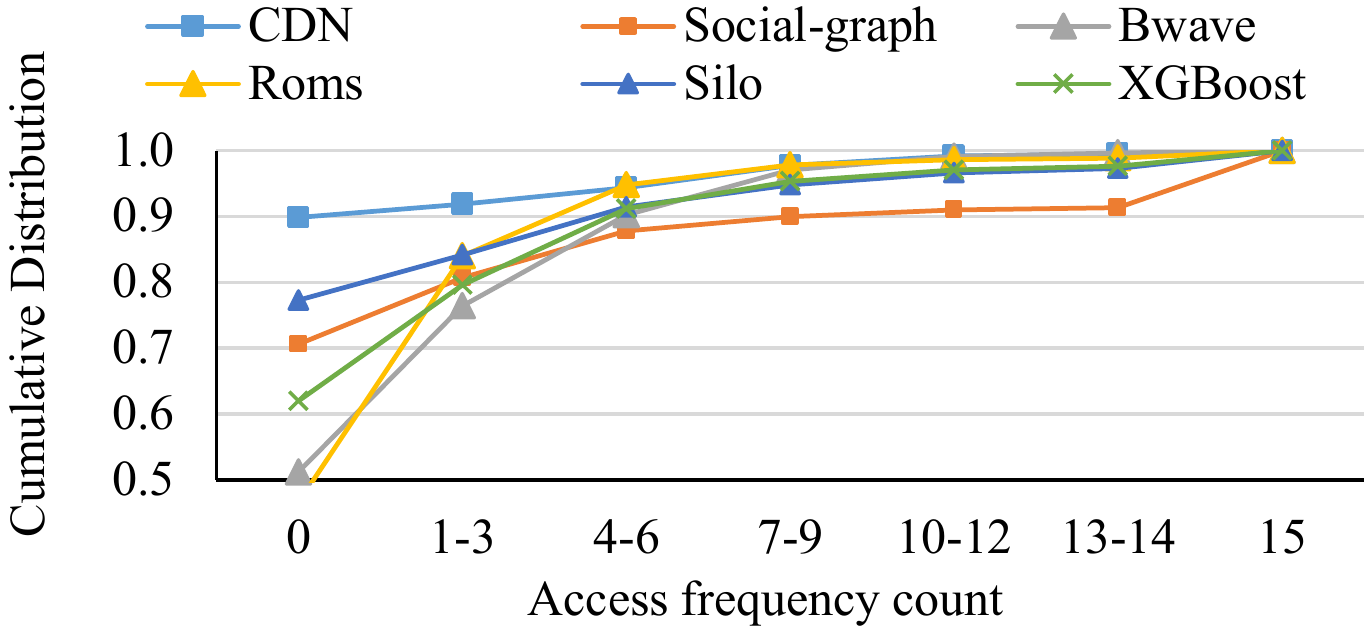}}
  \hspace{0.3mm}
  \subfloat[GAP. K: Kronecker graph, U: uniform random.]{\includegraphics[width=0.35\textwidth]{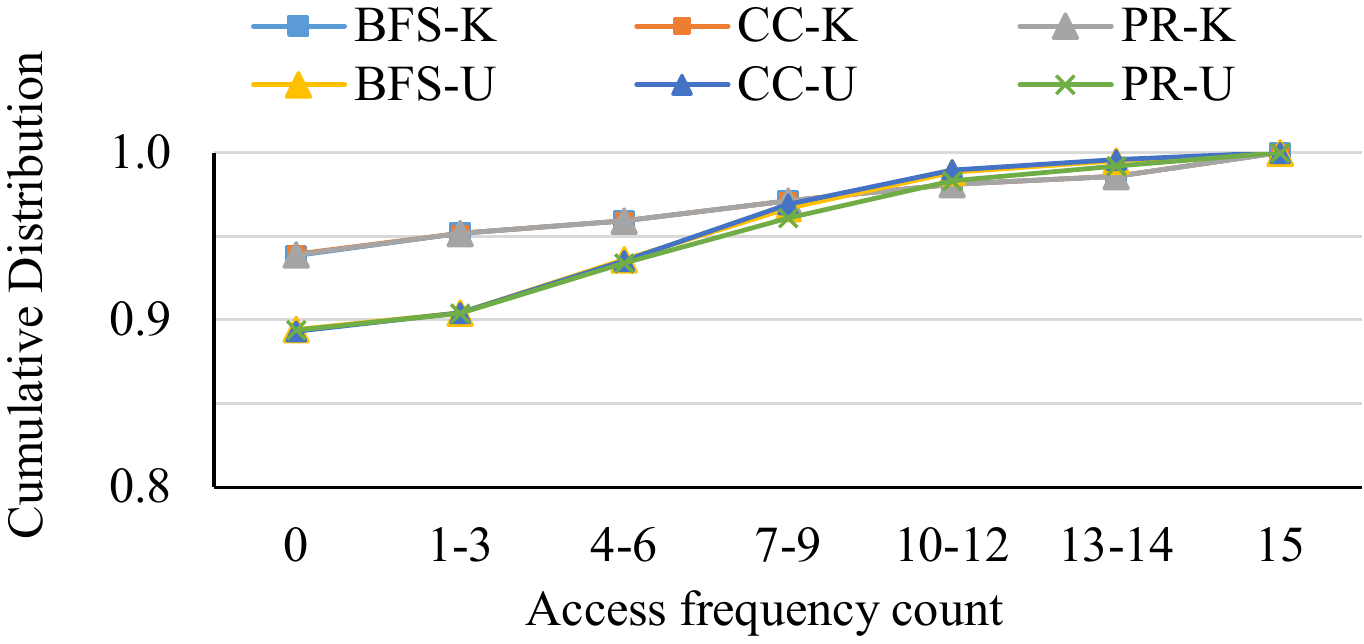}}
  \caption{Access hotness distributions of 12 workloads evaluated. Social-graph has the largest fraction of pages with access count >= 15, equal to ~25GB of memory.}
\label{fig:freq_dist}
\Description{}
\end{figure}

We first justify \projectname{}'s design decision to cap the size of each access counter to 4 bits.
From Figure \ref{fig:freq_dist}, we observe that for all workloads except for social-graph, the fraction of pages with frequency $\geq$ 15 is less than 3$\%$.
If the ratio between fast and slow-tier memory is greater than 3\% (or roughly 1:32), which is common in practical settings, such pages should all be placed in fast-tier memory.
Thus, the tiering system can treat such pages equally and does not need to differentiate between them. 
In cases where more than 4 bits are required, users can choose to either increase the counter width to 8 or 16 at the expense of higher memory and cache overhead.

Next, we wish to understand the accuracy of CBF.
To measure CBF accuracy, we modify \projectname{} to maintain a hash table in addition to the CBF.
%
%
Since the hash table guarantees exactness (Section \ref{sec:reducing_num_metadata}), we use access statistics in the hash table as the ground truth.
%
%
\autoref{table:cbf_accuracy} shows that a 64MB CBF agrees with the hash table for more than 99.6\% of migration decisions.
In the 1:16 configuration, \projectname{} uses 128MB CBF to ensure high tracking accuracy. 
This confirms that despite being probabilistic, CBF can introduce minimal inaccuracy in the context of memory tiering. 

\begin{table}[t]
    \small 
    \centering
    \caption{Accuracy of migration decisions made by counting bloom filter running CacheLib under 1:16 configuration.
     } 
    \vspace{-3mm}
    \setlength\tabcolsep{8pt}
    \aboverulesep = 0.2mm \belowrulesep = 0.2mm 
    \addtolength{\tabcolsep}{-0.25em}
    \begin{tabular}{c c c c c c} 
     \toprule
      CBF size (MB) & 256 & 128 & 64 & 32 & 8   \\
     \midrule
    Accuracy  &  99.72\% & 99.65\%	& 99.62\% & 99.42\% & 96.92\% \\ 
     \bottomrule
    \end{tabular}
     \vspace{-1mm}
    \label{table:cbf_accuracy}
\end{table}

\begin{figure}[t]
\centering
\includegraphics[width=0.35\textwidth]{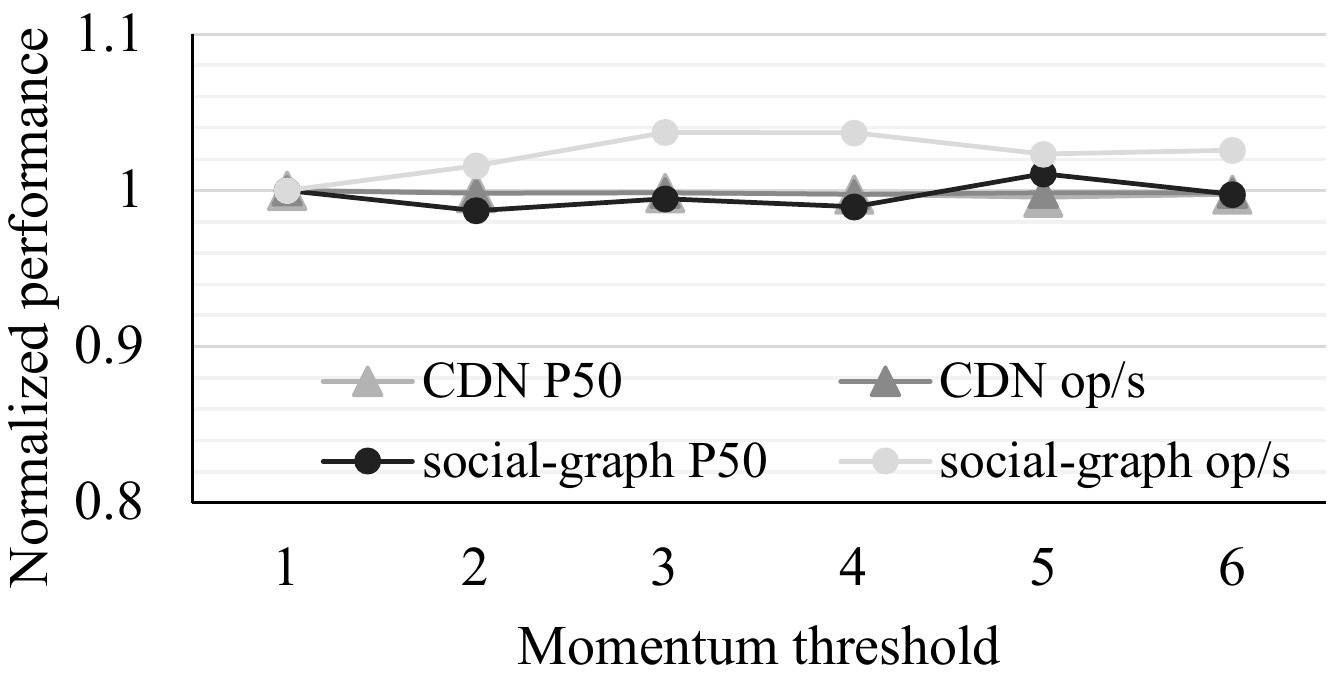}
\caption{Momentum threshold sensitivity.}
\label{fig:momentum_thresh}
\Description{}
\end{figure}

\subsubsection{Momentum Threshold} \label{sec:momentum_thresh}
We conduct a sensitivity study to understand the effect of momentum thresholds on \projectname{} performance, shown in Figure \ref{fig:momentum_thresh}. 
The social-graph workload is more sensitive to momentum threshold than CDN since it has a larger hot set (Figure \ref{fig:freq_dist}), making fast-tier memory more scarce. 
Reducing the momentum threshold below 3 negatively impacts \projectname{}'s performance, since a cold page that is accessed only a few times can be mistakenly promoted.
Increasing the momentum threshold beyond 3 does not significantly improve performance for workloads we evaluated.
If the default momentum threshold is suboptimal for a specific application, \projectname{} allows users to adjust its value to improve performance.

%% file: sec/07_discussion.tex
\section{Discussions}
\paragraph{Userspace vs. Kernel Tiering}
We implement \projectname{} in the userspace due to its flexibility.
This advantage is recognized by both prior memory tiering systems \cite{hemem, aifm} and software systems such as userspace networking systems \cite{snap, dpdk} and file systems \cite{fuse}.
\projectname{}’s core design principles (frequency-momentum metrics, probabilistic access tracking) are general and can be implemented in kernel space.

\paragraph{Global Tiering}
To support global memory tiering (e.g., multi-tenant VM, co-located applications), one could use a central \projectname{} controller that coordinates with individual \projectname{} instances. 
Each \projectname{} instance would report local hot/cold items to the central controller, which makes global promotion/demotion decisions.
\paragraph{One-time-only Access Patterns} 
%
To handle applications with one-time-only access patterns such as scanning and pointer chasing, the user may tune \projectname{}’s momentum hotness threshold parameter. 
A higher momentum threshold makes \projectname{} more resistant to fast-tier memory pollution due to one-time-only accesses, but at the same time reduces \projectname{}’s ability to adapt to hotness changes.
%


\paragraph{Selecting Configurable Parameters}
HybridTier automatically adjusts the frequency hotness threshold based on the hotness distribution, similar to the approach taken by Memtis. 
We empirically the momentum hotness threshold to 3, which works well across workloads we evaluated. We perform a sensitivity study in section \ref{sec:momentum_thresh}.

%% file: sec/08_related_works.tex
\section{Related Works} \label{sec:related_works}


\paragraph{Memory Tiering Systems.}
Tiered memory systems can utilize various types of memory technologies as the slower tier.
Examples include solid state drives \cite{cachelib_ssd, ssd_tiering1, ssd_tiering2} or persistent memory \cite{hemem, multiclock, thermostat, heteroos, google_nvm, nvm_keyvalstore}.  
The slow-tier memory can be placed on the same server as the fast-tier memory (near memory) \cite{hemem, tpp, autonuma_huang}, or on remote servers (far memory)~\cite{nimble, aifm, carbink}, where memory accesses are served from the network. 
In terms of programmability, tiering systems range from application-transparent \cite{linux_autonuma, tpp, hemem, zswap, tmo} to requiring manual modifications to the application \cite{aifm, memkind}. 

AutoNUMA \cite{autonuma_huang} and TPP \cite{tpp} are recency-based CXL tiering systems. Memtis \cite{memtis} is a frequency-based tiering system designed for both persistent memory and CXL memory. 
HeMem \cite{hemem} is a persistent memory tiering system that performs frequency-based tiering at the page granularity.  
Application-guided tiering systems such as memkind \cite{memkind}, SMDK \cite{smdk}, Unimem \cite{unimem}, Xmem \cite{xmem}, and 2PP \cite{2pp} offer a potential for tiering systems to understand application semantics through profiling or application modifications.  Prior works \cite{memory_mode, hw_tier1, hw_tier2} have accelerated tiering using specialized hardware. 
In contrast, \projectname{} does not require any changes to the hardware or application.
Memstrata~\cite{memstrata} is a multi-VM memory allocator that utilizes Intel Flat Memory Mode, which requires hardware support. \projectname{} is a runtime solution that is hardware agnostic. NOMAD~\cite{nomad} focuses on removing page migration from the critical path of program execution. Colloid~\cite{colloid} focuses on balancing access latencies. Both NOMAD and Colloid can be integrated with existing memory tiering systems such as \projectname{}.

\paragraph{General Caching Algorithms.}
LRU tracks item access recency. CLOCK \cite{clock} is an approximation of LRU and thus suffers from the same drawback \cite{arc}.
\projectname{} adopts a hybrid frequency-recency caching algorithm. Among prior hybrid caching algorithms, we found the following to be the most relevant:
LRFU \cite{lrfu} tracks the weighted average for each item. By adjusting the decay factor, LRFU provides a spectrum of policies between LRU and LFU. Memtis \cite{memtis} adopts a variation of LRFU policy with decay factor of 1/2. However, LRFU combines the access recency and frequency of an item into a single weighted average, LRFU cannot track both frequency and recency accurately. For example, a lower decay factor captures frequency more accurately but sacrifices recency. \projectname{} addresses this challenge by allocating two counters to track frequency and recency independently.
ARC \cite{arc} maintains two LRU lists, one for items seen only once, and the other for items seen at least twice. The first list estimates “recency” and the second “frequency”. This approach has been adopted by several memory tiering works, such as TPP \cite{tpp} and Multi-CLOCK \cite{multiclock}. However, since ARC does not maintain frequency counts \cite{arc}, it cannot distinguish between hot and warm items, which is critical for memory tiering in resource-constrained scenarios \cite{memtis}

\paragraph{Approximate Data Structures.} Caching policies using approximate data structures have been proposed in various domains \cite{prob_cache1, prob_cache2, prob_cache3, prob_cache4}. 
More specifically, prior works study counting bloom filters from the theory side \cite{theory_cbf1, theory_cbf2} or apply counting bloom filters in domains such as database, in-memory caching, and network communication \cite{hadoop_cbf, tinylfu, caffeine, theory_cbf3}. While we drew inspiration from these works, to our knowledge, \projectname{} is the first work to practically apply the unique advantages of CBF to memory tiering. 
\projectname{} introduces efficient promotion/demotion mechanisms based on CBF's unique properties, alongside flexible migration policies to make fine-grained frequency-based tiering practical.
\paragraph{Disaggregated Memory Systems.} Memory disaggregation \cite{carbink, aifm, nimble, leap} expands the main memory capacity by placing additional memory modules in remote servers. Similar to memory tiering on a single machine, disaggregated memory systems also aim to place the hottest data on the local fast-tier memory. Our work is orthogonal to works in this domain. In addition to locally-attached CXL memory managed by \projectname{}, the target server can further expand its memory capacity through disaggregated tiering systems.








%% file: sec/09_conclusion.tex
\section{Conclusion}


We propose \projectname{}, an adaptive and lightweight tiered memory system. \projectname{} quickly adapts to changing access distributions by tracking both long-term data access frequency and short-term access momentum simultaneously. At the same time, \projectname{} achieves low metadata memory and cache overhead by adopting probabilistic access frequency tracking.
\projectname{} outperforms prior works on a wide range of workloads and memory configurations while reducing tiering memory and cache overhead. 

%% file: sec/appendix.tex
\section{Artifact Appendix}

\subsection{Abstract}


We provide the source code of HybridTier and scripts to reproduce the results presented in Figures 9, 10, and 12. 
The artifact includes the HybridTier runtime and various open-source large memory workloads. 
To facilitate the AE process, we provide access to remote access to the authors' machine with the pre-installed software.

\subsection{Artifact check-list (meta-information)}


{\small
\begin{itemize}
  \item {\bf Algorithm:}  Memory tiering
  \item {\bf Compilation:}  g++
  \item {\bf Hardware:}  Multi-NUMA node systems
  \item {\bf Metrics:}  End-to-end wall clock time
  \item {\bf Experiments:}  End-to-end evaluation using regular page (4KB) and huge page (2MB)
  \item {\bf Experiments:}  Reproduce part of the results in Figures 9, 10, and 12
  \item {\bf How much disk space required (approximately)?: } 20GB
  \item {\bf How much time is needed to prepare workflow (approximately)?: } 5 minutes
  \item {\bf How much time is needed to complete experiments (approximately)?: } 45 hours
  \item {\bf Publicly available?: } Yes
  \item {\bf Code licenses (if publicly available)?:} MIT license
\end{itemize}
}

\subsection{Description}

\subsubsection{How to access}
The artifact is available for download on GitHub: \url{https://github.com/kevins981/hybridtier-asplos25-artifact}. To facilitate the AE process, we provide reviewers with remote server access. In such a case, please skip directly to section \ref{sec:flow}. Otherwise, we provide detailed instructions to reproduce in the GitHub repository.

\subsubsection{Hardware dependencies}
Requires a single x86\_64 Linux host. The processor must support Processor Event-Based Sampling (PEBS). The system must have multiple tiers of memory. The paper uses a system with two NUMA nodes to emulate tiered CXL memory systems. The provided source code will not work out of the box on a system with e.g., real CXL memory or persistent memory. To do so, the processor performance counters must be modified to capture the appropriate memory access events. We provide more details in the GitHub repository.

\subsubsection{Software dependencies}
This artifact depends on the following
environment.
\begin{itemize}
  \item Ubuntu 20.04.4 LTS
  \item g++ 9.4.0
\end{itemize}

While the HybridTier runtime does not require a specific kernel, emulating CXL memory using remote NUMA node requires a few minor changes to the Linux kernel. 
We provide details on the kernel modifications required and the source code of the kernel we used to evaluate HybridTier in the GitHub repository. 


\subsection{Installation}
\begin{lstlisting}[style=BashInputStyle]
git clone https://github.com/kevins981/hybridtier-asplos25-artifact.git
\end{lstlisting}
This artifact has the following structure:

{\small
\begin{itemize}
\item \texttt{repro.sh}: Reproduce major results
\item \texttt{run\_\{workload\}.sh}: Run experiments for a particular workload.
\item \texttt{tiering\_runtime/}: HybridTier implementation
\item \texttt{hook/}: source code for launching HybridTier runtime
\item \texttt{tools/}: auxiliary tools used for experiments
\end{itemize}
}

\subsection{Experiment workflow} \label{sec:flow}
We provide a script, \texttt{repro.sh}, to reproduce the major results.
Experiment progress can be found in \texttt{./exp\_log}.
After all experiments complete, \texttt{repro\_postprocess.sh} extracts results and create plots. 

\begin{lstlisting}[style=BashInputStyle]
cd hybridtier-asplos25-artifact
sudo ./repro.sh
# wait for all experiments (~45 hours)
./repro_postprocess.sh
\end{lstlisting}

To evaluate different tiering systems and fast:slow memory ratios, \texttt{repro.sh} will automatically load the appropriate kernel and reboot the server. Thus, it is normal for the server to be temporarily offline during reboots. 
Upon reboot, \texttt{repro.sh} will automatically perform the next experiments. No manual interventions are needed during this process.

Each experiment requires approximately 1 hour to complete. As a result, running all experiments in Figures 9, 10, and 12 would consume more than 200 hours. 
To reduce time  required for artifact evaluation, we made the following changes to the experiments in Figures 9, 10, and 12:
\begin{itemize}
  \item Out of the 6 tiering systems, we reproduce HybridTier (ours) and Memtis, the best-performing prior tiering system.
  \item Out of the 3 memory configurations, we reproduce 1:4 and 1:16 fast:slow memory ratios. 1:4 represents systems with an abundance of fast-tier memory, and 1:16 represents systems with limited fast-tier memory.
  \item Out of the two input graphs used for graph analytic workload (GAP), we reproduce results using the kron graph, which is closer to real-world graphs than uniform random graph.
  \item We leave out SPEC CPU experiments as the workload is not open source.
  \item We leave out XGBoost experiments as each run can take close to 2 hours to complete.
\end{itemize}

\subsection{Evaluation and expected results}
After all experiments are completed, \texttt{repro\_results.csv} will be created that summarizes the result of all experiments performed. This file contains both absolute and relative performance results.
Each table is also plotted and saved under \texttt{figs/}. 

%% file: main.bbl
\begin{thebibliography}{10}

\bibitem{thermostat}
Neha Agarwal and Thomas~F. Wenisch.
\newblock Thermostat: {Application}-transparent page management for two-tiered main memory.
\newblock In {\em Proceedings of the Twenty-Second International Conference on Architectural Support for Programming Languages and Operating Systems (ASPLOS)}, page 631–644, New York, NY, USA, 2017. Association for Computing Machinery.

\bibitem{xgboost_cpu1}
Amazon.
\newblock X{GB}oost algorithm.
\newblock \url{https://docs.aws.amazon.com/sagemaker/latest/dg/xgboost.html\#Instance-XGBoost-training-cpu}.
\newblock Accessed: 2024.

\bibitem{hadoop_cbf}
{Apache Software Foundation}.
\newblock Class countingbloomfilter.
\newblock \url{https://hadoop.apache.org/docs/r2.7.5/api/org/apache/hadoop/util/bloom/CountingBloomFilter.html}.
\newblock Accessed: 2024.

\bibitem{fuse}
The Linux~Kernel Archives.
\newblock Fuse.
\newblock \url{https://www.kernel.org/doc/html/next/filesystems/fuse.html}.
\newblock Accessed: 2024.

\bibitem{gap}
Scott Beamer, Krste Asanović, and David Patterson.
\newblock The {GAP} benchmark suite, 2015.

\bibitem{cachelib_paper}
Benjamin Berg, Daniel~S. Berger, Sara McAllister, Isaac Grosof, Sathya Gunasekar, Jimmy Lu, Michael Uhlar, Jim Carrig, Nathan Beckmann, Mor Harchol-Balter, and Gregory~R. Ganger.
\newblock The {CacheLib} caching engine: Design and experiences at scale.
\newblock In {\em 14th USENIX Symposium on Operating Systems Design and Implementation (OSDI)}, pages 753--768, November 2020.

\bibitem{speccpu}
James Bucek, Klaus-Dieter Lange, and J\'{o}akim v.~Kistowski.
\newblock {SPEC CPU2017}: Next-generation compute benchmark.
\newblock In {\em Companion of the 2018 ACM/SPEC International Conference on Performance Engineering}, ICPE '18, page 41–42, New York, NY, USA, 2018. Association for Computing Machinery.

\bibitem{cachebench}
CacheLib.
\newblock Cachebench overview.
\newblock \url{https://cachelib.org/docs/Cache_Library_User_Guides/Cachebench_Overview/}.
\newblock Accessed: 2024.

\bibitem{cachelib_ssd}
CacheLib.
\newblock Hybrid cache.
\newblock \url{https://cachelib.org/docs/Cache_Library_Architecture_Guide/hybrid_cache}.
\newblock Accessed: 2024.

\bibitem{ycsb}
Brian~F. Cooper, Adam Silberstein, Erwin Tam, Raghu Ramakrishnan, and Russell Sears.
\newblock Benchmarking cloud serving systems with ycsb.
\newblock In {\em Proceedings of the 1st ACM Symposium on Cloud Computing}, SoCC '10, page 143–154, New York, NY, USA, 2010. Association for Computing Machinery.

\bibitem{clock}
F.J. Corbat{\'o} and Project MAC (Massachusetts~Institute of~Technology).
\newblock {\em A Paging Experiment with the Multics System}.
\newblock Project MAC. Massachusetts Institute of Technology, 1968.

\bibitem{pagemetadata1}
Jonathan Corbet.
\newblock Cramming more into struct page.
\newblock \url{https://lwn.net/Articles/565097/}, 2013.

\bibitem{pagemetadata2}
Jonathan Corbet.
\newblock Persistent memory support progress.
\newblock \url{https://lwn.net/Articles/640113/}, 2015.

\bibitem{criteo_link}
Criteo.
\newblock Criteo {1TB} click logs dataset.
\newblock \url{https://ailab.criteo.com/criteo-1tb-click-logs-dataset/}.
\newblock Accessed: 2024.

\bibitem{dpdk}
DPDK.
\newblock Data plane development kit.
\newblock \url{https://github.com/DPDK/dpdk}.
\newblock Accessed: 2024.

\bibitem{prob_cache2}
Amit Dua, Megha Shishodia, Nikhil Kumar, Gagangeet~Singh Aujla, and Neeraj Kumar.
\newblock Bloom filter based efficient caching scheme for content distribution in vehicular networks.
\newblock In {\em 2019 IEEE International Conference on Communications Workshops (ICC Workshops)}, pages 1--6, 2019.

\bibitem{xmem}
Subramanya~R. Dulloor, Amitabha Roy, Zheguang Zhao, Narayanan Sundaram, Nadathur Satish, Rajesh Sankaran, Jeff Jackson, and Karsten Schwan.
\newblock Data tiering in heterogeneous memory systems.
\newblock In {\em Proceedings of the Eleventh European Conference on Computer Systems (EuroSys)}, New York, NY, USA, 2016. Association for Computing Machinery.

\bibitem{google_nvm}
Padmapriya Duraisamy, Wei Xu, Scott Hare, Ravi Rajwar, David Culler, Zhiyi Xu, Jianing Fan, Christopher Kennelly, Bill McCloskey, Danijela Mijailovic, Brian Morris, Chiranjit Mukherjee, Jingliang Ren, Greg Thelen, Paul Turner, Carlos Villavieja, Parthasarathy Ranganathan, and Amin Vahdat.
\newblock Towards an adaptable systems architecture for memory tiering at warehouse-scale.
\newblock In {\em Proceedings of the 28th ACM International Conference on Architectural Support for Programming Languages and Operating Systems (ASPLOS)}, page 727–741, New York, NY, USA, 2023. Association for Computing Machinery.

\bibitem{tinylfu}
Gil Einziger, Roy Friedman, and Ben Manes.
\newblock {TinyLFU: A} highly efficient cache admission policy.
\newblock {\em ACM Trans. Storage}, 13(4), nov 2017.

\bibitem{cachelib_git}
Facebook.
\newblock Cachelib.
\newblock \url{https://github.com/facebook/CacheLib}.
\newblock Accessed: 2024.

\bibitem{cbf_orig}
Li~Fan, Pei Cao, Jussara Almeida, and Andrei~Z. Broder.
\newblock Summary cache: {A} scalable wide-area web cache sharing protocol.
\newblock {\em SIGCOMM Comput. Commun. Rev.}, 28(4):254–265, oct 1998.

\bibitem{prob_cache4}
Jie Gao, Shan Zhang, Lian Zhao, and Xuemin Shen.
\newblock The design of dynamic probabilistic caching with time-varying content popularity.
\newblock {\em IEEE Transactions on Mobile Computing}, 20(4):1672--1684, 2021.

\bibitem{daemon}
Christina Giannoula, Kailong Huang, Jonathan Tang, Nectarios Koziris, Georgios Goumas, Zeshan Chishti, and Nandita Vijaykumar.
\newblock {DaeMon: Architectural Support for Efficient Data Movement in Fully Disaggregated Systems}.
\newblock {\em Proc. ACM Meas. Anal. Comput. Syst.}, 7(1), March 2023.

\bibitem{roms}
The~ROMS/TOMS Group.
\newblock 654.roms spec cpu 2017 benchmark description.
\newblock \url{https://www.spec.org/cpu2017/Docs/benchmarks/654.roms_s.html}.

\bibitem{ssd_tiering1}
Herodotos Herodotou and Elena Kakoulli.
\newblock Automating distributed tiered storage management in cluster computing.
\newblock {\em Proc. VLDB Endow.}, 13(1):43–56, sep 2019.

\bibitem{pagemetadata0}
Jian Huang, Moinuddin~K. Qureshi, and Karsten Schwan.
\newblock An evolutionary study of linux memory management for fun and profit.
\newblock In {\em 2016 USENIX Annual Technical Conference (USENIX ATC 16)}, pages 465--478, Denver, CO, June 2016.

\bibitem{hint_fault_patch}
Ying Huang.
\newblock memory tiering: hot page selection with hint page fault latency.
\newblock \url{https://patchwork.kernel.org/project/linux-mm/patch/20210722031819.3446711-5-ying.huang@intel.com/}.
\newblock Accessed: 2024.

\bibitem{ying_comment}
Ying Huang.
\newblock memory tiering: hot page selection with hint page fault latency.
\newblock \url{https://lore.kernel.org/linux-mm/bf23f05830db51bab3b06bac6e54d4743d37e955.camel@intel.com/}.
\newblock Accessed: 2024.

\bibitem{autonuma_huang}
Ying Huang.
\newblock [patch -v4 0/3] memory tiering: hot page selection.
\newblock \url{https://lwn.net/ml/linux-kernel/20220622083519.708236-1-ying.huang@intel.com/}.
\newblock Accessed: 2024.

\bibitem{bloom_calc}
Thomas Hurst.
\newblock Bloom filter calculator.
\newblock \url{https://hur.st/bloomfilter/}.
\newblock Accessed: 2024.

\bibitem{prob_cache1}
Hideo Inagaki, Ryota Kawashima, and Hiroshi Matsuo.
\newblock Improving apache spark's cache mechanism with lrc-based method using bloom filter.
\newblock In {\em 2018 Sixth International Symposium on Computing and Networking Workshops (CANDARW)}, pages 496--500, 2018.

\bibitem{xgboost_cpu2}
Intel.
\newblock Maximize your {CPU} resources for {XGBoost} training and inference.
\newblock \url{https://www.intel.com/content/www/us/en/developer/videos/maximize-cpu-resources-xgboost-training-inference.html#gs.47qye6}.
\newblock Accessed: 2024.

\bibitem{memory_mode}
Intel.
\newblock Why is the {Intel Optane Persistent Memory} in {Memory Mode} not persistent?
\newblock \url{https://www.intel.com/content/www/us/en/support/articles/000055895/memory-and-storage/intel-optane-persistent-memory.html}.
\newblock Accessed: 2024.

\bibitem{twoq}
Theodore Johnson and Dennis Shasha.
\newblock {2Q}: A low overhead high performance buffer management replacement algorithm.
\newblock In {\em Proceedings of the 20th International Conference on Very Large Data Bases}, VLDB '94, page 439–450, San Francisco, CA, USA, 1994. Morgan Kaufmann Publishers Inc.

\bibitem{ssd_tiering2}
Elena Kakoulli and Herodotos Herodotou.
\newblock {OctopusFS: A} distributed file system with tiered storage management.
\newblock In {\em Proceedings of the 2017 ACM International Conference on Management of Data (SIGMOD)}, page 65–78, New York, NY, USA, 2017. Association for Computing Machinery.

\bibitem{heteroos}
Sudarsun Kannan, Ada Gavrilovska, Vishal Gupta, and Karsten Schwan.
\newblock {HeteroOS - OS} design for heterogeneous memory management in datacenter.
\newblock In {\em ACM/IEEE 44th Annual International Symposium on Computer Architecture (ISCA)}, pages 521--534, 2017.

\bibitem{nvm_keyvalstore}
Hiwot~Tadese Kassa, Jason Akers, Mrinmoy Ghosh, Zhichao Cao, Vaibhav Gogte, and Ronald Dreslinski.
\newblock Improving performance of flash based {Key-Value} stores using storage class memory as a volatile memory extension.
\newblock In {\em USENIX Annual Technical Conference (ATC)}, pages 821--837, July 2021.

\bibitem{bwaves}
Mark Kremenetsky.
\newblock 603.bwaves {SPEC CPU} 2017 benchmark description.
\newblock \url{https://www.spec.org/cpu2017/Docs/benchmarks/603.bwaves_s.html}.
\newblock Accessed: 2024.

\bibitem{zswap}
Andres Lagar-Cavilla, Junwhan Ahn, Suleiman Souhlal, Neha Agarwal, Radoslaw Burny, Shakeel Butt, Jichuan Chang, Ashwin Chaugule, Nan Deng, Junaid Shahid, Greg Thelen, Kamil~Adam Yurtsever, Yu~Zhao, and Parthasarathy Ranganathan.
\newblock Software-defined far memory in warehouse-scale computers.
\newblock In {\em Proceedings of the Twenty-Fourth International Conference on Architectural Support for Programming Languages and Operating Systems (ASPLOS)}, page 317–330, New York, NY, USA, 2019. Association for Computing Machinery.

\bibitem{lrfu}
Donghee Lee, Jongmoo Choi, Jong-Hun Kim, S.H. Noh, Sang~Lyul Min, Yookun Cho, and Chong~Sang Kim.
\newblock {LRFU}: a spectrum of policies that subsumes the least recently used and least frequently used policies.
\newblock {\em IEEE Transactions on Computers}, 50(12):1352--1361, 2001.

\bibitem{memtis}
Taehyung Lee, Sumit~Kumar Monga, Changwoo Min, and Young~Ik Eom.
\newblock Memtis: Efficient memory tiering with dynamic page classification and page size determination.
\newblock In {\em Proceedings of the 29th Symposium on Operating Systems Principles}, SOSP '23, page 17–34, New York, NY, USA, 2023. Association for Computing Machinery.

\bibitem{pond}
Huaicheng Li, Daniel~S. Berger, Lisa Hsu, Daniel Ernst, Pantea Zardoshti, Stanko Novakovic, Monish Shah, Samir Rajadnya, Scott Lee, Ishwar Agarwal, Mark~D. Hill, Marcus Fontoura, and Ricardo Bianchini.
\newblock Pond: {CXL}-based memory pooling systems for cloud platforms.
\newblock In {\em Proceedings of the 28th ACM International Conference on Architectural Support for Programming Languages and Operating Systems (ASPLOS)}, page 574–587, New York, NY, USA, 2023. Association for Computing Machinery.

\bibitem{theory_cbf3}
Wenjing Liu, Zhiwei Xu, Jie Tian, and Yujun Zhang.
\newblock Towards in-network compact representation: {Mergeable} counting bloom filter vis cuckoo scheduling.
\newblock {\em IEEE Access}, PP:1--1, 04 2021.

\bibitem{caffeine}
Ben Manes.
\newblock Caffeine.
\newblock \url{https://github.com/ben-manes/caffeine}.
\newblock Accessed: 2024.

\bibitem{snap}
Michael Marty, Marc de~Kruijf, Jacob Adriaens, Christopher Alfeld, Sean Bauer, Carlo Contavalli, Mike Dalton, Nandita Dukkipati, William~C. Evans, Steve Gribble, Nicholas Kidd, Roman Kononov, Gautam Kumar, Carl Mauer, Emily Musick, Lena Olson, Mike Ryan, Erik Rubow, Kevin Springborn, Paul Turner, Valas Valancius, Xi~Wang, and Amin Vahdat.
\newblock Snap: a microkernel approach to host networking.
\newblock In {\em In ACM SIGOPS 27th Symposium on Operating Systems Principles (SOSP)}, 2019.

\bibitem{multiclock}
Adnan Maruf, Ashikee Ghosh, Janki Bhimani, Daniel Campello, Andy Rudoff, and Raju Rangaswami.
\newblock {MULTI-CLOCK: Dynamic} tiering for hybrid memory systems.
\newblock In {\em IEEE International Symposium on High-Performance Computer Architecture (HPCA)}, pages 925--937, 2022.

\bibitem{leap}
Hasan~Al Maruf and Mosharaf Chowdhury.
\newblock Effectively prefetching remote memory with leap.
\newblock In {\em USENIX Annual Technical Conference (ATC)}, pages 843--857, July 2020.

\bibitem{tpp}
Hasan~Al Maruf, Hao Wang, Abhishek Dhanotia, Johannes Weiner, Niket Agarwal, Pallab Bhattacharya, Chris Petersen, Mosharaf Chowdhury, Shobhit Kanaujia, and Prakash Chauhan.
\newblock {TPP: Transparent} page placement for {CXL}-enabled tiered-memory.
\newblock In {\em Proceedings of the 28th ACM International Conference on Architectural Support for Programming Languages and Operating Systems (ASPLOS)}, New York, NY, USA, 2023. Association for Computing Machinery.

\bibitem{arc}
Nimrod Megiddo and Dharmendra~S. Modha.
\newblock {ARC}: A {Self-Tuning}, low overhead replacement cache.
\newblock In {\em 2nd USENIX Conference on File and Storage Technologies (FAST 03)}, San Francisco, CA, March 2003.

\bibitem{hw_tier1}
Mitesh~R. Meswani, Sergey Blagodurov, David Roberts, John Slice, Mike Ignatowski, and Gabriel~H. Loh.
\newblock Heterogeneous memory architectures: A hw/sw approach for mixing die-stacked and off-package memories.
\newblock In {\em IEEE 21st International Symposium on High Performance Computer Architecture (HPCA)}, pages 126--136, 2015.

\bibitem{cxl_micron}
Micron.
\newblock {CZ120} memory expansion module.
\newblock \url{https://www.micron.com/solutions/server/cxl#:~:text=CXL%20memory%20expansion%20serves%20as,workloads%20for%20CXL%2Denabled%20servers.}
\newblock Accesed: 2024.

\bibitem{autonuma_gap}
Diego Moura, Daniel Mosse, and Vinicius Petrucci.
\newblock Performance characterization of {AutoNUMA} memory tiering on graph analytics.
\newblock In {\em {IEEE} International Symposium on Workload Characterization ({IISWC})}. {IEEE}, nov 2022.

\bibitem{mem_scaling_onur}
Onur Mutlu.
\newblock Memory scaling: A systems architecture perspective.
\newblock \url{https://users.ece.cmu.edu/~omutlu/pub/mutlu_memory-scaling_memcon13_talk.pdf}, 2013.

\bibitem{theory_cbf1}
Sabuzima Nayak and Ripon Patgiri.
\newblock {countBF}: {A} general-purpose high accuracy and space efficient counting bloom filter.
\newblock In {\em 17th International Conference on Network and Service Management (CNSM)}, 2021.

\bibitem{smdk}
OpenMPDK.
\newblock Scalable memory development kit.
\newblock \url{https://github.com/OpenMPDK/SMDK}.
\newblock Accessed: 2024.

\bibitem{cxl_pan}
Panmnesia.
\newblock Panmnesia technologies.
\newblock \url{https://panmnesia.com/#technology}.
\newblock Accessed: 2024.

\bibitem{mem_prospect}
J.~Thomas Pawlowski.
\newblock Prospects for memory.
\newblock \url{https://passlab.github.io/mchpc/mchpc2019/presentations/MCHPC_Pawlowski_keynote.pdf}.
\newblock Accessed: 2024.

\bibitem{aws_t2}
Amazon {EC2 T2} instances.
\newblock \url{https://aws.amazon.com/ec2/instance-types/t2/}.
\newblock Accessed: 2024.

\bibitem{ibs}
{AMD} research instruction based sampling toolkit.
\newblock \url{https://github.com/jlgreathouse/AMD_IBS_Toolkit}.
\newblock Accessed: 2024.

\bibitem{cxl}
{Computer Express Link}.
\newblock \url{https://computeexpresslink.org/}.
\newblock Accessed: 2024.

\bibitem{caffeine_bcbf}
A high performance caching library for java.
\newblock \url{https://github.com/ben-manes/caffeine/blob/3f4c1599992accac7d596e3047fcb0866cabe048/caffeine/src/main/java/com/github/benmanes/caffeine/cache/FrequencySketch.java#L42}.
\newblock Accessed: 2024.

\bibitem{memtis_pagetable}
Memtis: Efficient memory tiering with dynamic page classification and page size determination.
\newblock \url{https://github.com/cosmoss-jigu/memtis/blob/838a802680a8a760d3dea50754d6ea8a8530f6aa/linux/mm/htmm_core.c#L1030}.
\newblock Accesed: 2024.

\bibitem{bcbf}
Modern bloom filters: 22x faster!
\newblock \url{https://save-buffer.github.io/bloom_filter.html#org7b03738}.
\newblock Accessed: 2024.

\bibitem{windows_perf}
Performance counters tools.
\newblock \url{https://learn.microsoft.com/en-us/windows/win32/perfctrs/performance-counters-tools}.
\newblock Accessed: 2024.

\bibitem{tiering0.8}
Release tiering-0.8.
\newblock \url{https://kernel.googlesource.com/pub/scm/linux/kernel/git/vishal/tiering/+/refs/tags/tiering-0.8}.
\newblock Accessed: 2024.

\bibitem{uthash}
U{T} hash.
\newblock \url{https://troydhanson.github.io/uthash/}.
\newblock Accessed: 2024.

\bibitem{memkind}
{pmem.io}.
\newblock memkind.
\newblock \url{https://pmem.io/memkind/}.
\newblock Accessed: 2024.

\bibitem{hw_tier2}
Luiz~E. Ramos, Eugene Gorbatov, and Ricardo Bianchini.
\newblock Page placement in hybrid memory systems.
\newblock In {\em Proceedings of the International Conference on Supercomputing (ICS)}, page 85–95, New York, NY, USA, 2011. Association for Computing Machinery.

\bibitem{hemem}
Amanda Raybuck, Tim Stamler, Wei Zhang, Mattan Erez, and Simon Peter.
\newblock {HeMem: Scalable} tiered memory management for big data applications and real {NVM}.
\newblock In {\em Proceedings of the ACM SIGOPS 28th Symposium on Operating Systems Principles (SOSP)}, page 392–407, New York, NY, USA, 2021. Association for Computing Machinery.

\bibitem{mtm}
Jie Ren, Dong Xu, Junhee Ryu, Kwangsik Shin, Daewoo Kim, and Dong Li.
\newblock {MTM}: Rethinking memory profiling and migration for multi-tiered large memory.
\newblock In {\em Proceedings of the Nineteenth European Conference on Computer Systems}, EuroSys '24, page 803–817, New York, NY, USA, 2024. Association for Computing Machinery.

\bibitem{theory_cbf2}
Pedro Reviriego and Ori Rottenstreich.
\newblock The tandem counting bloom filter - it takes two counters to tango.
\newblock {\em IEEE/ACM Transactions on Networking}, 27(6):2252--2265, 2019.

\bibitem{linux_autonuma}
Vinod~Chegu Rik~van Riel.
\newblock Automatic numa balancing.
\newblock \url{https://www.linux-kvm.org/images/7/75/01x07b-NumaAutobalancing.pdf}, 2014.
\newblock Accesed: 2024.

\bibitem{aifm}
Zhenyuan Ruan, Malte Schwarzkopf, Marcos~K. Aguilera, and Adam Belay.
\newblock {AIFM}: {High-Performance}, {Application-Integrated} far memory.
\newblock In {\em 14th USENIX Symposium on Operating Systems Design and Implementation (OSDI)}, pages 315--332, November 2020.

\bibitem{cxl_samsung}
Samsung.
\newblock Samsung electronics introduces industry’s first 512gb {CXL} memory module.
\newblock \url{https://news.samsung.com/us/samsung-electronics-introduces-industrys-first-512gb-cxl-memory-module/}.
\newblock Accesed: 2024.

\bibitem{cxl_whitepaper1}
Debendra~Das Sharma.
\newblock Introduction to compute express link.
\newblock \url{https://docs.wixstatic.com/ugd/0c1418_d9878707bbb7427786b70c3c91d5fbd1.pdf}.
\newblock Accessed: 2024.

\bibitem{prob_cache3}
David Starobinski and David Tse.
\newblock Probabilistic methods for web caching.
\newblock {\em Perform. Eval.}, 46(2–3):125–137, October 2001.

\bibitem{uiuc_cxl}
Yan Sun, Yifan Yuan, Zeduo Yu, Reese Kuper, Chihun Song, Jinghan Huang, Houxiang Ji, Siddharth Agarwal, Jiaqi Lou, Ipoom Jeong, Ren Wang, Jung~H. Ahn, Tianyin Xu, and Kim~Nam S.
\newblock Demystifying cxl memory with genuine cxl-ready systems and devices.
\newblock In {\em 2023 56th IEEE/ACM International Symposium on Microarchitecture (MICRO)}, 2023.

\bibitem{twitter_data}
Sysomos.
\newblock Inside twitter: An in-depth look inside the {Twitter} world.
\newblock \url{https://www.key4biz.it/files/000270/00027033.pdf}, 2014.
\newblock Accesed: 2024.

\bibitem{silo}
Stephen Tu, Wenting Zheng, Eddie Kohler, Barbara Liskov, and Samuel Madden.
\newblock Speedy transactions in multicore in-memory databases.
\newblock In {\em Proceedings of the Twenty-Fourth ACM Symposium on Operating Systems Principles}, SOSP '13, page 18–32, New York, NY, USA, 2013. Association for Computing Machinery.

\bibitem{colloid}
Midhul Vuppalapati and Rachit Agarwal.
\newblock Tiered memory management: Access latency is the key!
\newblock In {\em Proceedings of the ACM SIGOPS 30th Symposium on Operating Systems Principles}, SOSP '24, page 79–94, New York, NY, USA, 2024. Association for Computing Machinery.

\bibitem{2pp}
Wei Wei, Dejun Jiang, Sally~A. McKee, Jin Xiong, and Mingyu Chen.
\newblock Exploiting program semantics to place data in hybrid memory.
\newblock In {\em International Conference on Parallel Architecture and Compilation (PACT)}, pages 163--173, 2015.

\bibitem{tmo}
Johannes Weiner, Niket Agarwal, Dan Schatzberg, Leon Yang, Hao Wang, Blaise Sanouillet, Bikash Sharma, Tejun Heo, Mayank Jain, Chunqiang Tang, and Dimitrios Skarlatos.
\newblock {TMO: Transparent} memory offloading in datacenters.
\newblock In {\em Proceedings of the 27th ACM International Conference on Architectural Support for Programming Languages and Operating Systems (ASPLOS)}, page 609–621, New York, NY, USA, 2022. Association for Computing Machinery.

\bibitem{unimem}
Kai Wu, Yingchao Huang, and Dong Li.
\newblock Unimem: {Runtime} data management on non-volatile memory-based heterogeneous main memory.
\newblock In {\em Proceedings of the International Conference for High Performance Computing, Networking, Storage and Analysis (SC)}, New York, NY, USA, 2017. Association for Computing Machinery.

\bibitem{xgboost_git}
XGBoost.
\newblock {XGBoost}: {eXtreme} gradient boosting.
\newblock \url{https://github.com/dmlc/xgboost}.
\newblock Accessed: 2024.

\bibitem{nomad}
Lingfeng Xiang, Zhen Lin, Weishu Deng, Hui Lu, Jia Rao, Yifan Yuan, and Ren Wang.
\newblock Nomad: non-exclusive memory tiering via transactional page migration.
\newblock In {\em Proceedings of the 18th USENIX Conference on Operating Systems Design and Implementation}, OSDI'24, USA, 2024.

\bibitem{flexmem}
Dong Xu, Junhee Ryu, Kwangsik Shin, Pengfei Su, and Dong Li.
\newblock {FlexMem}: Adaptive page profiling and migration for tiered memory.
\newblock In {\em 2024 USENIX Annual Technical Conference (USENIX ATC 24)}, pages 817--833, Santa Clara, CA, July 2024.

\bibitem{nimble}
Zi~Yan, Daniel Lustig, David Nellans, and Abhishek Bhattacharjee.
\newblock Nimble page management for tiered memory systems.
\newblock In {\em Proceedings of the Twenty-Fourth International Conference on Architectural Support for Programming Languages and Operating Systems (ASPLOS)}, page 331–345, New York, NY, USA, 2019. Association for Computing Machinery.

\bibitem{twitter_paper}
Juncheng Yang, Yao Yue, and K.~V. Rashmi.
\newblock A large scale analysis of hundreds of in-memory cache clusters at {Twitter}.
\newblock In {\em 14th USENIX Symposium on Operating Systems Design and Implementation (OSDI)}, pages 191--208, November 2020.

\bibitem{memstrata}
Yuhong Zhong, Daniel~S. Berger, Carl Waldspurger, Ryan Wee, Ishwar Agarwal, Rajat Agarwal, Frank Hady, Karthik Kumar, Mark~D. Hill, Mosharaf Chowdhury, and Asaf Cidon.
\newblock Managing memory tiers with {CXL} in virtualized environments.
\newblock In {\em 18th USENIX Symposium on Operating Systems Design and Implementation (OSDI 24)}, pages 37--56, Santa Clara, CA, July 2024.

\bibitem{carbink}
Yang Zhou, Hassan M.~G. Wassel, Sihang Liu, Jiaqi Gao, James Mickens, Minlan Yu, Chris Kennelly, Paul Turner, David~E. Culler, Henry~M. Levy, and Amin Vahdat.
\newblock Carbink: {Fault-tolerant} far memory.
\newblock In {\em 16th USENIX Symposium on Operating Systems Design and Implementation (OSDI)}, pages 55--71, Carlsbad, CA, July 2022.

\end{thebibliography}
